\documentclass[aps,pra,twocolumn,showpacs,floatfix]{revtex4}
\usepackage{epsfig}
\usepackage{graphicx}
\usepackage{dcolumn}
\usepackage{amsthm,amsmath}
\usepackage{rotating}
\usepackage{pdflscape}
\usepackage{longtable}

\begin{document}
\title{Annexing magic and tune-out wavelengths to the clock transitions of the alkaline-earth metal ions}

\author{Jasmeet Kaur\footnote{Email: jasmeetphy.rsh@gndu.ac.in}, Sukhjit Singh and Bindiya Arora\footnote{Email: bindiya.phy@gndu.ac.in}}
\affiliation{Department of Physics, Guru Nanak Dev University, Amritsar, Punjab-143005, India}

\author{B. K. Sahoo\footnote{Email: bijaya@prl.res.in}}
\affiliation{Atomic, Molecular and Optical Physics Division, Physical Research Laboratory, Navrangpura, Ahmedabad-380009, India}

\date{Received date; Accepted date}

\begin{abstract}
We present additional magic wavelengths ($\lambda_{\rm{magic}}$) for the clock transitions in the alkaline-earth metal ions 
considering circular polarized light aside from our previously reported values in [J. Kaur et al., Phys. Rev. A {\bf 92}, 031402(R) (2015)]
for the linearly polarized light. Contributions from 
the vector component to the dynamic dipole polarizabilities ($\alpha_d(\omega)$) of the atomic states associated with the clock 
transitions play major roles in the evaluation of these $\lambda_{\rm{magic}}$, hence facilitating in choosing circular polarization 
of lasers in the experiments. Moreover, the actual clock transitions in these ions are carried out among the hyperfine levels. The
$\lambda_{\rm{magic}}$ values in these hyperfine transitions are estimated and found to be different from $\lambda_{\rm{magic}}$ for 
the atomic transitions due to different contributions coming from the vector and tensor part of $\alpha_d(\omega)$. Importantly, we also 
present $\lambda_{\rm{magic}}$ values that depend only on the scalar component of $\alpha_d(\omega)$ for their uses in a specially designed trap 
geometry for these ions so that they can be used unambiguously among any hyperfine levels of the atomic states of the clock transitions. 
We also present $\alpha_d(\omega)$ values explicitly at the 1064 nm for the atomic states associated with the clock transitions which may 
be useful for creating ``high-field seeking'' traps for the above ions using the Nd:YAG laser. The tune out wavelengths at which the states 
would be free from the Stark shifts are also presented. Accurate values of the electric dipole matrix elements required for these studies 
are given and trends of electron correlation effects in determining them are also highlighted. 
\end{abstract}

\maketitle

\section{Introduction}\label{introsec}

Atomic clocks based on optical lattices are capable of proffering outstanding stable and accurate time keeping devices. A 
fundamental feature of an optical lattice clock is that it interrogates an optical transition with the controlled atomic motion 
\cite{Poli}. At present, the most stable clock is based on the optical lattices of $^{87}$Sr atoms with accuracy below $10^{-18}$ \cite{Nicholoson}. 
The unique feature of this clock is that the atoms are trapped at the wavelengths of an external electric field at which the differential
light shift of an atomic transition nullifies. These wavelengths are commonly known as the magic wavelengths ($\lambda_{\rm{magic}}$) 
\cite{katori}. However, ions provide more accurate atomic clocks since various systematics in the ions can be controlled better 
~\cite{Champenois,Chou}. As a result a number of ions, such as, $^{27}$Al$^+$ $^{199}$Hg$^+$, $^{171}$Yb$^+$, $^{115}$In$^+$, $^{88}$Sr$^+$,
$^{40}$Ca$^+$, $^{113}$Cd$^+$ etc. are under consideration for building accurate clocks.  Among the various ions proposed 
for frequency standards~\cite{margolis,Peik,Rosenband,Stalnaker,Champenois}, the alkaline earth metal ions possess an advantage that 
transitions required for cooling and re-pumping of ions and clock frequency measurement can be easily accessed using non-bulky solid state or
diode lasers \cite{Champenois}. Moreover presence of metastable $D$ states in these ions, whose lifetimes range from milliseconds to several 
seconds, assist in carrying out measurements meticulously. 

The recent development on measurement of $\lambda_{\rm{magic}}$ in a singly charged $^{40}$Ca$^+$ ion has now open-up a platform for 
the possibility of building up pure optical trapped ion clocks \cite{Liu}. Optical lattices blended with unique features of optical 
transitions in an ionic system can revolutionize the secondary as well as primary frequency standards. However, the potential of the optical 
dipole trap perturb the energy levels of the ion unevenly and the consistency of an 
ion optical clock depreciates. Therefore, knowledge of $\lambda_{\rm{magic}}$ values in these ions would be instrumental for constructing 
all-optical trapped ion clocks. These wavelengths can be found out using accurate values of the dynamic dipole polarizabilities $\alpha_d(\omega)$
of the states associated with the clock transitions. Also, information on the dynamic ($\alpha_d(\omega)$) values, especially at which the ions are 
being trapped will be of great significance. Improved atomic clocks will obviously ease the widely used technologies including precise determination of 
fundamental constants \cite{book1}, accurate control of quantum states \cite{Sackett} and advancement in communication, Global 
Positioning System~\cite{Hong} etc. 

Following the measurement of $\lambda_{\rm{magic}}$ in the $4S_{1/2} - 3D_{5/2}$ transition of $^{43}$Ca$^+$, we had investigated 
these values in the $nS_{1/2} - (n-1)D_{3/2}$ and $nS_{1/2} - (n-1)D_{5/2}$ transitions of the alkaline-earth ions \cite{jasmeet2} for the ground state principal quantum 
number $n$. However, those studies were focused mainly on the linearly polarized light limiting the choice for experimental measurements.
Application of circular polarized light to atomic systems introduce contributions from the vector polarizabilities in the Stark shifts 
of the energy levels, which is linearly proportional to the angular frequency of the applied field. This can help in manipulating the 
Stark shifts in the energy levels and can lend to more degrees of freedom to attain further $\lambda_{\rm{magic}}$ values as per the 
experimental stipulation. Moreover, it is advantageous to consider hyperfine transitions in certain isotopes of the singly charged 
alkaline-earth ions, giving rise to zero hyperfine angular momentum, to get-rid off the systematics due to electric 
quadrupole shifts~\cite{Sahoo09,Sherman,Sahoo07}. 
Since $\alpha_d$ values of atomic and hyperfine states in an atomic system are different when vector and tensor components of $\alpha_d$ 
contribute, the $\lambda_{\rm{magic}}$ values also differ between the atomic and hyperfine transitions. Thus, it would be pertinent 
to investigate $\lambda_{\rm{magic}}$ both in the atomic and hyperfine transitions before the experimental consideration of the 
proposed ions. In fact, it could be more convenient to have $\lambda_{\rm{magic}}$ that are independent of choice of both magnetic  
and hyperfine sub-levels in a given clock transition.

In fact, the Nd:YAG lasers at 1064 nm is often used for trapping atoms and ions because of their relatively high power and 
low intensity noise \cite{Burda}. Traps built with long wavelength lasers are generally “high-field-seeking” where the atoms are attracted to the intensity
maxima. Dynamic polarizabilities for the considered ions at 1064 nm will be of immense interest to the 
experimentalists since these polarizabilities will be immediately useful for operating optical traps at 1064 nm light fields.

\begin{table*}
\caption{\label{pol1} Calculated values of the static dipole polarizabilities $\alpha^J_{d,i}(0)$  for $nS_{1/2}$ and $(n-1)D_{3/2,5/2}$ states of $^{43}$Ca$^+$, $^{87}$Sr$^+$ and $^{137}$Ba$^+$ alkaline earth metal ions. Polarizability 
values are compared with the other available theoretical and experimental results. References are given in square
brackets. Uncertainties in the results are given in parentheses.}
\begin{ruledtabular}
\begin{tabular}{lccccccc}
%Contribution & $\alpha_{0}$ & Contribution & $\alpha_{0}$ & $\alpha_{2}$ &  Contribution & $\alpha_{0}$ & $\alpha_{2}$\\
\multicolumn{8}{c}{Ca$^+$}  \\
\multicolumn{2}{c}{$4S_{1/2}$ state}                         & \multicolumn{3}{c}{$3D_{3/2}$ state}    &  \multicolumn{3}{c}{$3D_{5/2}$ state} \\
    & $\alpha_{d,0}^J$ &   & $\alpha_{d,0}^J$ & $\alpha_{d,2}^J$ &   & $\alpha_{d,0}^J$ & $\alpha_{d,2}^J$\\
\cline{1-2} \cline{3-5} \cline{6-8} \\
 $4P_{1/2}$  & 24.38(6)      & $4P_{1/2}$        & 19.24(5)  & $-$19.24(5)        & $4P_{3/2}$  & 22.77(8)  & $-$22.77(8)   \\
 $5P_{1/2}$  &  0.07           &  $(5-7)P_{1/2}$       & 0.011     & $-$0.011     & $(5-7)P_{3/2}$  & 0.016 & $-$0.016 \\  
 $(6-7)P_{1/2}$  &   0.01          & $ 4P_{3/2}$       & 3.76      & 3.009 	  & $4F_{5/2}$  & 0.12      &  0.14   \\
 $4P_{3/2}$  &  48.4(17)       & $(5-7)P_{3/2}$        &  0.0027   & 0.002        & $6F_{5/2}$  & 0.056     & 0.06  \\
 $5P_{3/2}$  &  0.01           &  $ 4F_{5/2}$          & 2.49(1)   &  $-$0.49      & $4F_{5/2}$  & 2.39(14)  & $-$0.85  \\
 $(6-7)P_{3/2}$  &  0.019          &  $5F_{5/2}$       & 0.81      &  $-$0.16      & $5F_{7/2}$  & 0.76      &  $-$0.27\\
            &                 &  $6F_{5/2}$        & 0.36      &  $-$0.07      & $6F_{7/2}$  & 0.001     & $-$0.0005\\
 ``Main"$\alpha_{d,i}^{J,v}$ &     72.88     &                    & 26.67     &   $-$16.96    &             & 26.12  & $-$23.71\\
 $\alpha_{d,i}^{J,c}$                & 3.25               & & 3.25      &            &             &  3.25     &  \\
 ``Tail "$\alpha_{d,i}^{J,v}$             & 5.86$\times 10^{-2}$& & 3.75     &  $-$0.75      &             &  3.75     & $-$1.08 \\
 $\alpha_{d,i}^{J,cv}$               &$-$8.85$\times 10^{-2}$ &&$-$7.94$\times 10^{-3}$ &  &            & $-$1.02$\times 10^{-2}$ &  \\
 $\alpha_{d,i}^J(0)$(Present)              &76.1(2)   & & 33.67(1.8) & $-$17.71    &            & 33.11(1.8) & $-$24.78(4)  \\
 $\alpha_{d,i}^J(0)$(Other) \cite{Tang}    &75.28     & & 32.99     &  $-$17.88     &            & 32.81      &   $-$25.16      \\
  $\alpha_{d,i}^J(0)$(Other) \cite{sahooca09}& 73.0(1.5) & & 28.5(1.0)&  $-$15.8(7)   &            & 29.5(1.0)  &   $-$22.45(5)\\ 
 $\alpha_{d,i}^J(0)$(Other) \cite{mitroyca}&75.49     & & 32.73     &  $-$17.64     &            & 32.73      &  $-$25.20 \\
 $\alpha_{d,i}^J(0)$(Expt.) \cite{edward}  & 75.3(4)  &            &             &    &\\
 \hline
 \multicolumn{8}{c}{Sr$^+$}  \\
\multicolumn{2}{c}{$5S_{1/2}$ state}                               & \multicolumn{3}{c}{$4D_{3/2}$ state}    &  \multicolumn{3}{c}{$4D_{5/2}$ state} \\  
    & $\alpha_{d,0}^J$ &   & $\alpha_{d,0}^J$ & $\alpha_{d,2}^J$ &   & $\alpha_{d,0}^J$ & $\alpha_{d,2}^J$\\
\cline{1-2} \cline{3-5} \cline{6-8} \\
 $5P_{1/2}$  &  29.47(10)      & $5P_{1/2}$          & 38.67(2)   & $-$38.67(2)  & $5P_{3/2}$  & 44.16(3)  & $-$44.16(3)   \\
 $6P_{1/2}$  &  0.0008         &  $(6-8)P_{1/2}$         & 0.0063     & $-$0.0063    & $(6-8)P_{3/2}$  & 0.015     & $-$0.015 \\  
 $(7-8)P_{1/2}$  &   0.071         & $5P_{3/2}$          & 7.024(2)   & 5.62(2)    & $4F_{5/2}$  & 0.33      &  0.37  \\
 $5P_{3/2}$  &  56.93(27)      & $(6-8)P_{3/2}$          &  0.003    & 0.002      & $6F_{5/2}$  & 0.12     & 0.13  \\
 $6P_{3/2}$  &  0.0015         &$4F_{5/2}$           & 6.63(7)   &  $-$1.33      & $4F_{5/2}$  & 6.51(5)  & $-$2.32(2)  \\
 $(7-8)P_{3/2}$  &  0.0058         &$5F_{5/2}$           & 1.76      &  $-$0.35      & $5F_{7/2}$  & 1.72      &  $-$0.61\\
            &                 & $6F_{5/2}$           & 0.72      &  $-$0.14      & $6F_{7/2}$  & 0.69     & $-$0.25\\
 ``Main"$\alpha_{d,i}^{J,v}$ &     86.47     &                      & 54.81  &   $-$34.88    &             & 53.55  & $-$46.85\\
 $\alpha_{d,i}^{J,c}$                & 4.98                 & & 4.98     &         &             &  4.98     &  \\
 ``Tail"$\alpha_{d,i}^{J,v}$             & 1.96$\times 10^{-2}$& & 4.94      &  $-$1.0       &             &  4.96     & $-$1.44 \\
 $\alpha_{d,i}^{J,cv}$               &$-$0.19                 &&$-$1.77$\times 10^{-3}$ &              &            & $-$2.78$\times 10^{-2}$ &  \\
 $\alpha_{d,i}^J(0)$(Present)              &91.23(0.3)& & 64.7(2.5) & $-$35.88(5) &              & 63.5(2.5) & $-$48.29(7)  \\
 $\alpha_{d,i}^J(0)$(Other) \cite{SR}      & 89.99    & &  61.77     &    -   &               & 61.77     &   -       \\
 $\alpha_{d,i}^J(0)$(Other) \cite{sahoo}&88.29(1.0)   & &  61.43(52) &  35.42 &               &  62.87(75)&  $-$48.83(30)   \\
 $\alpha_{d,i}^J(0)$(Expt.) \cite{Barklem}  & 93.3(9)  & &   57.0    &   -    &               & 57.0      & - \\
 \hline   
 \multicolumn{8}{c}{Ba$^+$}  \\ 
\multicolumn{2}{c}{$6S_{1/2}$ state}      & \multicolumn{3}{c}{$5D_{3/2}$ state}    &  \multicolumn{3}{c}{$5D_{5/2}$ state} \\  
    & $\alpha_{d,0}^J$ &   & $\alpha_{d,0}^J$ & $\alpha_{d,2}^J$ &   & $\alpha_{d,0}^J$ & $\alpha_{d,2}^J$\\
\cline{1-2} \cline{3-5} \cline{6-8} \\
 $6P_{1/2}$  &  40.23(14)       & $6P_{1/2}$        & 22.51(13)  & $-$22.51(13)& $6P_{3/2}$  & 25.66(1)  & $-$25.66(1)   \\
 $7P_{1/2}$  &  0.005           &  $(7-8)P_{1/2}$       & 0.075     & $-$0.074     & $(7-8)P_{3/2}$  & 0.13     &$-$0.13 \\  
 $8P_{1/2}$  &   0.009          & $6P_{3/2}$       & 3.86      & 3.095 	     & $4F_{5/2}$  & 0.668(5)  &  0.763(2)   \\
 $6P_{3/2}$  &  73.93(4)       & $(7-8)P_{3/2}$        &  0.023   & 0.018        & $(5-6)F_{5/2}$  & 0.0955(1) & 0.109(2)  \\
 $7P_{3/2}$  &  0.011          &  $4F_{5/2}$      & 11.85(3)   &  $-$2.37      & $4F_{7/2}$  & 13.34(13)  & $-$4.76(5)  \\
 $8P_{3/2}$  &  0.013          &  $(5-6)F_{5/2}$       & 1.76      &  $-$0.35      & $5F_{7/2}$  & 2.863(34)      &  $-$1.022(17)\\
 %           &                 &  $nF_{5/2}$        & 0.36      &  -0.07      & $nF_{7/2}$  & 0.001     & -0.0005\\
 ``Main"$\alpha_{d,i}^{J,v}$ &     114.19    &                    & 40.07& $-$22.19    &             & 42.76  & $-$30.71\\
 $\alpha_{d,i}^{J,c}$                & 9.35               & & 9.35      &           &                &  9.35     &  \\
``Tail"$\alpha_{d,i}^{J,v}$             & 1.66$\times 10^{-2}$& & 4.75     &  $-$1.0      &                &  4.79     & $-$1.46 \\
 $\alpha_{d,i}^{J,cv}$               &$-$0.37                 &&$-$2.34$\times 10^{-2}$ &  &              &  $-$3.87$\times 10^{-2}$ &  \\
 $\alpha_{d,i}^J(0)$(Present)               & 123.7(5)   & & 54.17(2.5) & $-$22.19(4)  &            & 56.87(2.4) & $-$32.17(3)  \\
 $\alpha_{d,i}^J(0)$(Other) \cite{sahoo}    & 124.26(1.0)     & & 48.81(46)    &  $-$24.62(28)     &            & 50.67(58)     &   $-$30.85(31)      \\
 $\alpha_{d,i}^J(0)$(Other) \cite{snow}     &123.88(5)\\
 \end{tabular}
\end{ruledtabular}
\end{table*}

In the present work, we aim to search for the $\lambda_{\rm{magic}}$ for the $nS_{1/2} - (n-1)D_{{3/2},{5/2}}$ optical clock transitions, 
both in the atomic and hyperfine levels, in $^{43}$Ca$^+$ (with nuclear spin $I$=7/2), $^{87}$Sr$^+$ ($I$=9/2) and $^{137}$Ba$^+$ ($I$=3/2)
ions using circularly polarized light. These values can be compared with the values for the linearly polarized light reported in Ref. 
\cite{jasmeet2} for the experimental consideration to trap the above ions. Also, we had demonstrated in a recent work how a trap geometry can be chosen in such a 
way that Stark shifts observed by the energy levels can be free from the contributions from the vector and tensor components of the 
$\alpha_d$ of the atomic states \cite{sukhjitnew}. Assuming such trapping geometries for the considered alkaline-earth ions, we also give 
$\lambda_{\rm{magic}}$ values using only the scalar polarizability contributions. Moreover, we identify the tune-out wavelengths 
($\lambda_{\rm{T}}$) of the respective states for which the dynamic dipole polarizability of these ions vanishes. Comprehension of these 
$\lambda_{\rm{T}}$ values are needed for the sympathetic cooling of other possible singly and multiply charged ions in two-species 
mixtures with the considered alkaline earth ions \cite{Rosenband,Chou,Hobein}. We also present dynamic polarizability of these states 
at the 1064 nm wavelength of the applied external electric field. Contributions from various electric dipole (E1) matrix elements in
determining the $\alpha_d$ values and the role of the electron correlation effects for the evaluation of accurate values of the E1 matrix elements 
are also discussed. Unless stated otherwise, all the results are given in atomic unit (a.u.) throughout this paper.

\section{Theory}\label{thesec}

The Stark shift in the energy of $K^{th}$ level of an atom placed in an electric field is given by \cite{bonin,manakov}
\begin{equation}
\label{eq3}
\Delta E^{K}= -\frac{1}{4}\alpha_d^K (\omega){\cal E}^2,
\end{equation}
where ${\cal E}$ is the amplitude of the external electric field due to the applied laser in an atomic system and $\alpha_d^K(\omega)$ is the dynamic 
dipole polarizability for the state $K$ with its magnetic projection $M$. In tensor decomposition, $\alpha_d^K$ can be expressed as
\begin{eqnarray}
\alpha_d^K(\omega)&=&\alpha_{d,0}^K(\omega)+\beta(\epsilon)\frac{M}{2K}\alpha_{d,1}^K(\omega ) \nonumber \\ 
&& + \gamma(\epsilon) \frac{3M^2-K(K+1)}{K(2K-1)}\alpha_{d,2}^K(\omega),\label{eqmag} 
\end{eqnarray}
where $\alpha_{d,i}^K(\omega)$ with $i=0,1,2$ are the scalar, vector and tensor components of $\alpha_d^K(\omega)$ respectively. 
In the specific cases, $K$ can be either the atomic angular momentum $J$ or hyperfine 
angular momentum $F$. The terms $\beta(\epsilon)$ and $\gamma(\epsilon)$ are defined as~\cite{beloyt}
\begin{equation}
\beta(\epsilon) = \iota(\hat{\epsilon}\times\hat{\epsilon}^*)\cdot\hat{e}_B\label{beta}
\end{equation}
and
\begin{equation}
\gamma(\epsilon)= \frac{1}{2}[3(\hat{\epsilon}^*\cdot\hat{e}_B)(\hat{\epsilon}\cdot\hat{e}_B)-1], \label{gamma}
\end{equation}
with the quantization axis unit vector $\hat{e}_B$ and polarization unit vector $\hat{e}$. The differential Stark shift of a transition between 
states $K$ to $K'$ can be formulated as  
\begin{eqnarray}
\delta E_{KK'} &=& \Delta E_K - \Delta E_{K'} = -\frac{1}{2} \big [ \big\{\alpha_{d,0}^K ( \omega)-\alpha_{d,0}^{K'}(\omega) \big \} \nonumber \\ && 
+ \beta(\epsilon) \big \{ \frac{M_K}{2K}\alpha_{d,1}^K(\omega)  - \frac{M_{K'}}{2K'}\alpha_{d,1}^{K'}(\omega) \big \}  \nonumber \\ 
&& + \gamma(\epsilon) \big \{ \frac{3M_K^2-K(K+1)}{K(2K-1)}\alpha_{d,2}^K (\omega) \nonumber \\ 
&& -  \frac{3M_{K'}^2-K'(K'+1)}{K'(2K'-1)}\alpha_{d,2}^{K'} (\omega) \big \} \big ] {\cal E}^2, \label{eqmag1} 
\end{eqnarray}

To obtain null differential Stark shift, it is obvious from Eq. (\ref{eqmag1}) that either the independent components of the polarizabilities
cancel out each other or the net resultant nullifies which depend upon the choice of $\beta(\epsilon)$, $\gamma(\epsilon)$, $M_K$ and $M'_{K'}$ 
magnetic sublevels. Moreover, as we have demonstrated recently, the differential Stark shift can be independent of the vector and tensor 
components of the states involved in a transition for a certain trap geometry \cite{sukhjitnew}. Such trapping scheme is useful 
for $M_J$ , $F$ and $M_F$ insensitive trapping and can be suitably applied for considered clock transitions in alkaline earth metal ions.

\begin{table*}[t]\fontsize{7.5}{9.5}\selectfont
\caption{\label{pol2} Calculated values of the dynamic dipole polarizabilities $\alpha_{d,i}^J(\omega)$ for the $nS_{1/2}$ and $(n-1)D_{3/2,5/2}$ states 
of $^{43}$Ca$^+$, $^{87}$Sr$^+$ and $^{137}$Ba$^+$ ions at $\lambda$=1064 nm (equivalent to $\omega \simeq 0.035$ a.u.). 
Uncertainties in the results are given in parentheses.}
%\rotatebox[origin=c]{270}{
\begin{ruledtabular}
\begin{tabular}{lccccccccccc}
\multicolumn{11}{c}{Ca$^+$}  \\
\multicolumn{3}{c}{$4S$ state}                               & \multicolumn{4}{c}{$3D_{3/2}$ state}    &  \multicolumn{4}{c}{$3D_{5/2}$ state} \\ 
 & $\alpha_{d,0}^J$ & $\alpha_{d,1}^J$     &  & $\alpha_{d,0}^J$ &$\alpha_{d,1}^J$ &$\alpha_{d,2}^J$ &   & $\alpha_{d,0}^J$ &$\alpha_{d,1}^J$& $\alpha_{d,2}^J$\\
 \cline{2-3} \cline{5-7} \cline{9-11}\\
 $4P_{1/2}$  &  28.33(7)    &$-$21.141(5)  & $4P_{1/2}$     & 57.12(2) &$-$139.55(3)   & $-$57.12(2) & $4P_{3/2}$      & 64.13(2)  &$-$154.55(5) & $-$64.13(2)   \\
 $5P_{1/2}$  &  0.0072      &$-$0.002   & $(5-7)P_{1/2}$ & 0.011    &$-$0.007    & $-$0.011    & $(5-7)P_{3/2}$  & 0.016     & $-$0.009 & $-$0.016 \\  
 $(6-7)P_{1/2}$  & 0.011    &$-$0.0027  & $ 4P_{3/2}$    & 8.59(1)  &$-$8.24     & 6.87        & $4F_{5/2}$      & 0.12      & $-$0.018 &  0.14   \\
 $4P_{3/2}$  &  56.03(20)   & 20.719    & $(5-7)P_{3/2}$ & 0.0028   &$-$0.0006   & 0.0020      & $(5-6)F_{5/2}$  & 0.056     & $-$0.006 & 0.06  \\
 $5P_{3/2}$  &  0.01        & 0.002     & $ 4F_{5/2}$    & 2.57(1)  & 0.79       &  $-$0.51    & $4F_{5/2}$      & 2.46(12)  &  0.91    & $-$0.88  \\
 $(6-7)P_{3/2}$  &  0.019   & 0.003     & $5F_{5/2}$     & 0.82     & 0.22       &  $-$0.16    & $5F_{7/2}$      & 0.78      &  0.24   &  $-$0.27\\
            &               &           &$6F_{5/2}$      & 0.37     & 0.09       &  $-$0.07    & $6F_{7/2}$      & 0.001     &  0.0004 & $-$0.0005\\
 ``Main"$\alpha_{d,i}^{J,v}$      & 84.4    &$-$0.423  &        &69.5      & $-$146.69  &  $-$50.9    &                 & 67.5      &  $-$153.37 & $-$65.1\\
 $\alpha_{d,i}^{J,c}$      & 3.25    &          &        & 3.25     &            &             &                 &  3.25     &             & \\
 ``Tail"$\alpha_{d,i}^{J,v}$   & 5.90$\times 10^{-2}$& $-$3.49$\times 10^{-4}$ &&3.79 & 0.73 & $-$0.76&              &  3.80     & 0.81        & $-$1.09 \\
$\alpha_{d,i}^{J,cv}$      &$-$8.85$\times 10^{-2}$ &&&$-$7.94$\times 10^{-3}$ &  &&                             & $-$1.02$\times 10^{-2}$ & &  \\
 $\alpha_{d,i}^J(\omega)$  & 87.6(2)& $-$0.423   & & 76.5(1.9)  &$-$145.96(0.36)       & $-$51.7(3)  &                 & 74.6(1.9) & $-$152.56(0.41) & $-$66.0(5)   \\
\hline
 & & \\
 \multicolumn{11}{c}{Sr$^+$}  \\
\multicolumn{3}{c}{$5S$ state}                               & \multicolumn{4}{c}{$4D_{3/2}$ state}    &  \multicolumn{4}{c}{$4D_{5/2}$ state} \\    
  & $\alpha_{d,0}^J$ & $\alpha_{d,1}^J$     &  & $\alpha_{d,0}^J$ &$\alpha_{d,1}^J$ &$\alpha_{d,2}^J$ &   & $\alpha_{d,0}^J$ &$\alpha_{d,1}^J$& $\alpha_{d,2}^J$\\
 \cline{2-3} \cline{5-7} \cline{9-11}\\
 $5P_{1/2}$  &  34.96(12) & $-$27.714(9) & $5P_{1/2}$   & $-$730.9(4) & 2250.05(1.3)   & 730.9(4)  & $5P_{3/2}$     & 769.4(5)   & $-$2241.08(1.6)  & $-$769.4(5)   \\
 $6P_{1/2}$  &  0.0008    & $-$0.0003 &  $(6-8)P_{1/2}$ & 0.007       & $-$0.005       & $-$0.007  & $(6-8)P_{3/2}$ & 0.016      & $-$0.01          & $-$0.016 \\  
 $(7-8)P_{1/2}$  & 0.007  & -0.002    & $5P_{3/2}$      & 64.03(2)    & $-$72.49(2)    & 51.22(2)  & $4F_{5/2}$     & 0.34       & $-$0.059         &  0.38  \\
 $5P_{3/2}$  &  66.74(32) & 25.58(1)  & $(6-8)P_{3/2}$  &  0.003      & $-$0.0008      & 0.002     & $(5-6)F_{5/2}$ & 0.12       & $-$0.02          & 0.5  \\
 $6P_{3/2}$  &  0.0046    & 0.0003    &$4F_{5/2}$       & 6.91(7)     & 2.52           &  $-$1.38  & $4F_{5/2}$     & 6.79(5)    & 2.96             & $-$2.42(2)  \\
 $(7-8)P_{3/2}$  &  0.0029 & 0.0008   &$5F_{5/2}$       & 1.81        & 0.54           &  $-$0.36  & $5F_{7/2}$     & 1.76       & 0.63             &  $-$0.63\\
            &               &         & $6F_{5/2}$      & 0.73        & 0.22           &  $-$0.15  & $6F_{7/2}$    & 0.71       & 0.23             & $-$0.25\\
 ``Main"$\alpha_{d,i}^{J,v}$ & 101.7&$-$2.13 &          & $-$657.4    & 2180.82        &   780.2   &                & 779.1      &$-$2237.34        & $-$771.8\\
 $\alpha_{d,i}^{J,c}$     & 4.98      &       &         & 4.98        &           &           &              &  4.98     &   &  \\
 ``Tail"$\alpha_{d,i}^{J,v}$  & 1.96$\times 10^{-2}$&$-$8.13$\times 10^{-4}$ & & 5.05 &1.11  &  $-$1.02&          &  5.05    & 1.24   & $-$1.46 \\
 $\alpha_{d,i}^{J,cv}$    &$-$0.19 & & &$-$1.77$\times 10^{-3}$ & &             &   & $-$2.79$\times 10^{-2}$ & && \\
 $\alpha_{d,i}^J(\omega)$ &106.52(3) & -2.14 &   & $-$647.4(3)  &  2181.92(1.79) & 779.2(6) & &789.2(2.6)&$-$2236.10(2.82) & $-$773.(9)  \\
 \hline 
 & & \\
 \multicolumn{11}{c}{Ba$^+$}  &\\ 
\multicolumn{3}{c}{$6S$ state}      & \multicolumn{4}{c}{$5D_{3/2}$ state}    &  \multicolumn{4}{c}{$5D_{5/2}$ state} \\   
  & $\alpha_{d,0}^J$ & $\alpha_{d,1}^J$     &  & $\alpha_{d,0}^J$ &$\alpha_{d,1}^J$ &$\alpha_{d,2}^J$ &   & $\alpha_{d,0}^J$ &$\alpha_{d,1}^J$& $\alpha_{d,2}^J$\\
 \cline{2-3} \cline{5-7} \cline{9-11}\\
 $6P_{1/2}$  & 51.26(18) & $-$47.55(2) & $6P_{1/2}$        & 35.89(2) & $-$65.77(4) & $-$35.89(2)& $6P_{3/2}$    & 38.5(2)  &$-$66.08(3) & $-$38.5(2)   \\
 $7P_{1/2}$  &  0.006   &  $-$0.002 &  $(7-8)P_{1/2}$   & 0.077    & $-$0.05       & $-$0.077   & $(7-8)P_{3/2}$ & 0.13     & $-$0.08 & $-$0.13 \\  
 $8P_{1/2}$  &   0.009  &  $-$0.003 & $6P_{3/2}$       & 5.55      & $-$3.66(2)    & 4.44       & $4F_{5/2}$     & 0.70     & $-$0.13  &  0.80   \\
 $6P_{3/2}$  &  90.52(4)& 38.75(2)  & $(7-8)P_{3/2}$    &  0.024   & $-$-0.005     & 0.019      & $(5-6)F_{5/2}$  & 0.107(1) & $-$0.02  & 0.122(2)  \\
 $7P_{3/2}$  &  0.011   & 0.002     &  $4F_{5/2}$       & 12.44(3) & 4.85          &  $-$2.48   & $4F_{7/2}$      & 14.02(14)& 6.59     & $-$5.007(5)  \\
 $8P_{3/2}$  &  0.013   & 0.002     &  $(5-6)F_{5/2}$       & 1.83     & 0.5      &  $-$0.36 & $5F_{7/2}$      & 3.41(2) &1.29   &  $-$1.22\\
 %           &                 &  $nF_{5/2}$        & 0.36      &  -0.07      & $nF_{7/2}$  & 0.001     & -0.0005\\
 ``Main"$\alpha_{d,i}^{J,v}$ &141.81 & $-$8.80   &             &  55.8   &  $-$64.05 & $-$34.35  &             & 56.86     &$-$59.02 & $-$43.9\\
 $\alpha_{d,i}^{J,c}$  & 9.35 &           & & 9.35      &         &    &       &  9.35     &  \\
 ``Tail"$\alpha_{d,i}^{J,v}$   & 1.67$\times 10^{-2}$& $-$1.24$\times 10^{-4}$&& 3.29 &  0.7  &   $-$0.71      &             &  3.31 & 0.77   & $-$1.04 \\
$\alpha_{d,i}^{J,cv}$   &$-$0.37  &     &&$-$2.35$\times 10^{-2}$ &  &       &     &  $-$3.88$\times 10^{-2}$ &  \\
 $\alpha_{d,i}^J(\omega)$ & 150.6(5)& $-$8.80 & & 68.44(1.7)&$-$63.55(0.35) & $-$35.05(4)  &         & 69.52(1.7)& $-$58.25(0.38) & $-$44.97(5) \\
 \end{tabular}
\end{ruledtabular}
%}
\end{table*}

Conveniently the expressions for polarizabilities of the hyperfine and atomic levels can be expressed as \cite{sukhjitnew} 
\begin{eqnarray}
\alpha_d^{F}(\omega) &=& \alpha_{d,0}^{F} (\omega) + \alpha_{d,1}^{F} (\omega ) \frac{A \cos\theta_k M_{F}}{2F}
+ \alpha_{d,2}^{F}(\omega ) \nonumber \\
& \times &  \left( \frac{3\cos^2\theta_p-1}{2} \right)
 \left [ \frac{3M_{F}^2-F(F+1)}{F(2F-1)} \right ] , \label{totalF} \ \ \ \ \
\end{eqnarray}
and 
\begin{eqnarray}
\alpha_d^{J}(\omega) &=& \alpha_{d,0}^{J} (\omega) + \alpha_{d,1}^{J} (\omega ) \frac{A \cos\theta_k M_{J}}{2J}
+ \alpha_{d,2}^{J}(\omega ) \nonumber \\
& \times &  \left( \frac{3\cos^2\theta_p-1}{2} \right)
 \left [ \frac{3M_{J}^2-J(J+1)}{J(2J-1)} \right ] , \label{totalJ} \ \ \ \ \
\end{eqnarray}
where $\alpha_{d,i}^{K=J,F}(\omega)$ with $i=0,1,2$ are the scalar, vector and tensor components of the respective polarizabilities,
$A$ represents degree of polarization, $\theta_k$ is the angle between the quantization axis and wave vector, and $\theta_p$ is the angle 
between the quantization axis and direction of polarization of the field. In the presence of magnetic field, $\cos\theta_k$ and 
$\cos^2\theta_p$ can have any values depending on the direction of applied magnetic field. In the absence of magnetic field, 
$\cos\theta_k = 0$ and $\cos^2\theta_p = 1$ for the linearly polarized light, where polarization vector is assumed to be along the 
quantization axis. However it yields $\cos\theta_k = 1$ and $\cos^2\theta_p = 0$ for the circularly polarized light, where wave vector is 
assumed to be along the quantization axis. Polarizabilities of the hyperfine states can be evaluated from the atomic state results 
using the relations \cite{beloyt,dzubaflam}
\begin{eqnarray}
\alpha_{d,0}^{F}(\omega)&=&\alpha_{d,0}^{J}(\omega), \\
\alpha_{d,1}^{F}(\omega)&=&(-1)^{J+F+I+1}  \left\{ \begin{array}{ccc}
                             F & J & I\\
                          J & F & 1
                         \end{array}\right\}   \nonumber \\ &\times & \sqrt{\frac{F(2F+1)(2J+1)(J+1)}{J(F+1)}}
              \alpha_{d,1}^{J}(\omega) \ \
\end{eqnarray}
and
\begin{eqnarray}
\alpha_{d,2}^{F}(\omega)&=&(-1)^{J+F+I}  \left\{ \begin{array}{ccc}
                                            F & J & I\\
                                            J & F & 2
                                           \end{array}\right\}  \nonumber \\
                                           & \times & \sqrt{\frac{F (2F-1)(2F+1)}{(2F+3)(F+1)}} \nonumber \\
& \times & \sqrt{\frac{(2J+3)(2J+1)(J+1)}{J(2J-1)}}  \alpha_{d,2}^{J}(\omega) .
\end{eqnarray}

Further, we can evaluate the atomic dipole polarizabilities using the expressions
\begin{eqnarray}
\alpha_{d,0}^{J}(\omega)&=&\frac{2}{3(2J+1)} \sum_{J'} \frac{(E -E')|\langle J ||{\bf D}||J' \rangle|^2} {\omega^2 - (E -E')^2}, 
\label{scalar} \\
\alpha_{d,1}^{J}(\omega)&=& \sqrt{\frac{24J}{(J+1)(2J+1)}}\sum_{J'}(-1)^{J+J'}  \nonumber \\
              & & \times \left\{ \begin{array}{ccc}
                             J & 1 & J \\
                          1 & J' &1
                         \end{array}\right\} \frac{ \omega |\langle J ||{\bf D}|| J' \rangle|^2}  {\omega^2 - (E -E')^2 }  \label{vector}
\end{eqnarray}
and
\begin{eqnarray}
\alpha_{d,2}^{J}(\omega) &=& \sqrt{\frac{40 J (2J-1)}{3(J +1)(2J+3)(2J+1)}}
 \sum_{J'}(-1)^{J+J'} \nonumber \\ &\times &
                                  \left\{ \begin{array}{ccc}
                                            J & 2 & J \\
                                            1 & J' &1
                                           \end{array}\right\} \frac{ (E -E')|\langle J||{\bf D}||J' \rangle|^2} {\omega^2 - (E -E')^2}, \label{tensor}
\end{eqnarray}
where $J'$s are the angular momentum of the intermediate states, $E$ and $E'$ are energies of the corresponding states,
$\langle J||{\bf D}||J'\rangle$ are the reduced E1 matrix elements.

\begin{table*}[t!]\fontsize{8.0}{10.0}\selectfont
\caption{\label{magicd321} The $\lambda_{\rm{magic}}$ values (in nm) with their corresponding polarizabilities $\alpha^J_d(\lambda_{\rm{magic}})$ 
(in a.u.) for the $nS_{1/2}(M_{J}={1/2}) \rightarrow (n-1)D_{3/2}$ transitions using circularly polarized light in the $^{43}$Ca$^+$,$^{87}$Sr$^+$ and $^{137}$Ba$^+$ ions. The resonant wavelengths $\lambda_{\rm{res}}$ to indicate the placement of the $\lambda_{\rm{magic}}$ in between resonant transitions are also listed.}
\begin{tabular}{lcccccccccccc}
\hline
\hline
   && \multicolumn{2}{c}{$M_{J}$=3/2} & \multicolumn{2}{c}{$M_{J}$=1/2} & \multicolumn{2}{c}{$M_{J}$=$-$1/2} &\multicolumn{2}{c}{$M_{J}$=$-$3/2}  \\
\hline
Resonances & $\lambda_{\rm{res}}$ & $\lambda_{\rm{magic}}$ &$\alpha^J_d(\lambda_{\rm{magic}})$ &  $\lambda_{\rm{magic}}$ &$\alpha^J_d(\lambda_{\rm{magic}})$ &  $\lambda_{\rm{magic}}$ &$\alpha^J_d(\lambda_{\rm{magic}})$ &  $\lambda_{\rm{magic}}$ &$\alpha^J_d(\lambda_{\rm{magic}})$ \\
\hline\\
$^{43}$Ca$^+$  & \multicolumn{1}{c}{\underline{$4S_{1/2}-3D_{3/2}$}}   \\  
$4S_{1/2}-4P_{3/2}$ & 393.366 \\
		       &          & 394.6(4)  & $-20.34$ & 394.6(9) & $-2.82$ & 394.6(10) & 11.68 & 394.6(10) & 22.40 \\
$4S_{1/2}-4P_{1/2}$ & 396.847 & \\	
$3D_{3/2}-4P_{3/2}$ & 849.802 &\\
                    &             & 850.9(2) & 96.33 & 853.1(2) &  96.20  &      - &  - & - & - \\
$3D_{3/2}-4P_{1/2}$ & 866.214 &\\
	            &             & 1467.8(4) & 81.31  & 1013.4(5)  & 89.31 & 870.7(3) & 95.15 &- &-\\
$^{87}$Sr$^+$  & \multicolumn{1}{c}{\underline{$5S_{1/2}-4D_{3/2}$}}   \\ 
$5S_{1/2}-5P_{3/2}$ & 407.7714 \\
		      &      & 412.5(9) & $-20.62$ & 412.4(9) & 3.99 & 412.4(9)& 27.18 & 412.3(8) & 48.55 \\
$5S_{1/2}-5P_{1/2}$ & 421.5524 \\
$4D_{3/2}-5P_{3/2}$ & 1003.6654\\
                      &     & 1009.3(2) & 109.74 & 1019.7(3) & 109.29 & 1062.5(3) & 107.65 & - &- \\
$4D_{3/2}-5P_{1/2}$ & 1091.4874 & \\
                     &      &     -     &   -        & 1577.2(3) & 98.17 &-   & -  & - & -\\
$^{137}$Ba$^+$  & \multicolumn{1}{c}{\underline{$6S_{1/2}-5D_{3/2}$}} \\
$6S_{1/2}-6P_{3/2}$ & 455.4033 \\
		     &    & 468.8(4) & $-70.91$ & 468.1(4) & $-13.13$ & 467.6(5) & 30.14& 467.3(4) & 59.27 \\
$6S_{1/2}-6P_{1/2}$ & 493.4077 \\	
$5D_{3/2}-6P_{3/2}$ & 585.3675 \\
                    &      & 587.6(9)  & 373.89 & 589.5(3) & 368.60 & 589.6(5) & 368.35 & -& -\\ 
$5D_{3/2}-6P_{1/2}$ & 649.6898 \\
		      &  & 841.7(5) & 182.54  & 690.7(7) & 237.64 &  -  & - & - & - \\
\hline\\		      
\end{tabular}                
\end{table*}

\begin{table*}[t!]\fontsize{8.0}{10.0}\selectfont
\caption{\label{magicd322} The $\lambda_{\rm{magic}}$ values (in nm) with their corresponding polarizabilities $\alpha^J_d(\lambda_{\rm{magic}})$ 
(in a.u.) for the $nS_{1/2}(M_{J}=-{1/2}) \rightarrow (n-1)D_{3/2}$ transitions using circularly polarized light in the $^{43}$Ca$^+$,$^{87}$Sr$^+$ and $^{137}$Ba$^+$ ions. The resonant wavelengths $\lambda_{\rm{res}}$ to indicate the placement of the $\lambda_{\rm{magic}}$ in between resonant transitions are also listed.}
\begin{tabular}{lcccccccccccc}
\hline
\hline
   && \multicolumn{2}{c}{$M_{J}$=3/2} & \multicolumn{2}{c}{$M_{J}$=1/2} & \multicolumn{2}{c}{$M_{J}$=$-$1/2} &\multicolumn{2}{c}{$M_{J}$=$-$3/2}  \\
\hline
Resonances & $\lambda_{\rm{res}}$ & $\lambda_{\rm{magic}}$ &$\alpha^J_d(\lambda_{\rm{magic}})$ &  $\lambda_{\rm{magic}}$ &$\alpha^J_d(\lambda_{\rm{magic}})$ &  $\lambda_{\rm{magic}}$ &$\alpha^J_d(\lambda_{\rm{magic}})$ &  $\lambda_{\rm{magic}}$ &$\alpha^J_d(\lambda_{\rm{magic}})$ \\
\hline\\
$^{43}$Ca$^+$  & \multicolumn{1}{c}{\underline{$4S_{1/2}-3D_{3/2}$}}   \\  
%$4S_{1/2}-4P_{3/2}$ & 393.366 \\
%		       &          & 394.6(4)  & $-20.34$ & 394.6(9) & $-2.82$ & 394.6(10) & 11.68 & 394.6(10) & 22.40 \\
%$4S_{1/2}-4P_{1/2}$ & 396.847 & \\	
$3D_{3/2}-4P_{3/2}$ & 849.802 &\\
                    &             & 851.2(3) & 95.63 & 853.5(4) &  95.49  &      -      &   -     & - & - \\
$3D_{3/2}-4P_{1/2}$ & 866.214 &\\
	            &             & 1549.9(5) & 80.95 & 1031.4(4)  & 88.28 &  875.56(2) &  94.27 &- &-\\
$^{87}$Sr$^+$  & \multicolumn{1}{c}{\underline{$5S_{1/2}-4D_{3/2}$}}   \\ 
$5S_{1/2}-5P_{3/2}$ & 407.7714 \\
		      &      &  416.9(7) & $-$21.20  & 416.8(5) & 3.93 & 416.8(8)& 27.12 & 416.7(8) & 48.44 \\
$5S_{1/2}-5P_{1/2}$ & 421.5524 \\
                    &        & 442.5(5)& $-$25.36 & 442.3(6) & 1.94  & 442.1(2)& 26.46 & 441.9(4)& 48.03\\
         	     &       & 478.2(5)& $-$31.49 & 479.2(4) & $-$0.96 & 480.1(2)& 25.56 & 481.1(1)& 47.83\\
$4D_{3/2}-5P_{3/2}$ & 1003.6654\\
                      &      & 1009.2(1) & 106.69 & 1019.8(6) & 106.29 & 1064.8(5) & 104.79 & - &- \\
$4D_{3/2}-5P_{1/2}$ & 1091.4874 & \\
                     &      & & -& - 1593.1(5) & 96.54 &-   & -  & - & -\\
$^{137}$Ba$^+$  & \multicolumn{1}{c}{\underline{$6S_{1/2}-5D_{3/2}$}} \\
$6S_{1/2}-6P_{3/2}$ & 455.4033 \\
		     &    & 479.2(5) & $-79.42$ & 478.5(4) & $-17.21$ & 477.9(1) & 28.76 & 477.5(2) & 58.89 \\
$6S_{1/2}-6P_{1/2}$ & 493.4077 \\
		     &    & 505.4(1) & $-106.86$ & 505.1(4) & $-30.87$ & 504.8(2) & 24.05 & 504.7(2) & 57.93 \\
		     &    & 571.4(4) & $-314.04$ & 574.1(2) & $-213.32$ & 577.3(4) & $-125.29$ & - & - \\
$5D_{3/2}-6P_{3/2}$ & 585.3675 \\
                    &      & 589.4(4)  & 47.82 & 595.5(2) & 89.44 & 602.4(2) & 120.417 & 590.5(2)& 56.46\\ 
$5D_{3/2}-6P_{1/2}$ & 649.6898 \\
		      &  & 899.9(5) & 153.13  & 710.5(4) & 176.48   &  -  & - & - & - \\
\hline\\		      
\end{tabular}                
\end{table*}

We define the tune-out wavelength $\lambda_{\rm{T}}$ as the wavelength
where the dynamic polarizability of the state is zero. 
We have determined the tune out wavelengths for the ground, $(n-1)D_{3/2}$ and $(n-1)D_{5/2}$ states 
of $^{43}$Ca$^+$, $^{87}$Sr$^+$ and $^{137}$Ba$^+$ ions. The detailed description about these calculations 
have been given in the Refs. \cite{arora2011,Leblanc2007}.

\section{Method of evaluation}

As discussed in our earlier works \cite{nandy,recent,jasmeet2}, each component of $\alpha_d^J(\omega)$ can be conveniently evaluated in
the considered alkaline earth ions, in which many of the low-lying states have electronic configurations as a common closed core of inert 
gas atoms and a well defined valence orbital, by classifying into three different contributions such as
\begin{eqnarray}
\alpha_{d,i}^J(\omega)=\alpha_{d,i}^{J,c}(\omega)+\alpha_{d,i}^{J,cv}(\omega)+\alpha_{d,i}^{J,v}(\omega), 
\end{eqnarray}
where $\alpha_{d,i}^{J,c}$, $\alpha_{d,i}^{J,cv}$ and $\alpha_{d,i}^{J,v}$ are referred to as the core, core-valence and valence 
electron correlation contributions, respectively. Since the valence electron correlation effects are mainly estimated by 
$\alpha_{d,i}^{J,v}$, this gives majority contribution followed by $\alpha_{d,i}^{J,c}$. Again, accuracies in the {\it ab initio} 
results of these quantities mainly suffer due to the uncertainties associated with the calculated energies of the atomic states.
Therefore, we calculate only the E1 matrix elements of as many as low-lying states of the considered ions employing a relativistic 
coupled-cluster (RCC) method, which is an all order perturbative method, and combine them with the experimental energy values from 
the National Institute of Science and Technology (NIST) database \cite{NIST} to determine the dominant contributions to $\alpha_{d,i}^{J,v}$.
We refer this as ``Main'' contribution to $\alpha_{d,i}^{J,v}$; while the smaller contributions coming from the high-lying excited states are
estimated in the {\it ab initio} formalism using the Dirac-Hartree-Fock (DHF) method and are mentioned as ``Tail'' contribution to
$\alpha_{d,i}^{J,v}$. To estimate the other two contributions, it is not possible to use sum-over-states approach, so we determine 
$\alpha_{d,i}^{J,c}$ in the random-phase approximation (RPA). This included core-polarization effects to all orders. It has been found 
that RPA can give these values reliably for the inert gas configured atomic systems \cite{yashpal}. Again as demonstrated later, 
the $\alpha_{d,i}^{J,cv}$ contributions are very small in these ions. Thus, we evaluate them using the DHF method.
 
A small number of E1 matrix elements of the $nD-nF$ transitions of $^{137}$Ba$^+$ are borrowed from the work of Sahoo \textit{et al.} 
\cite{SahooBa}, while the rest E1 matrix elements for the evaluation of ``Main'' contribution to $\alpha_d^{J,v}$ are obtained considering 
the singles and doubles excitation approximation in the RCC (SD) method as described in Refs. \cite{Blundell,Johnson87}. In the SD 
method, the wave function of the state with the closed-core with a valence electron $v$ is represented 
as an expansion
\begin{eqnarray}
|\Psi_v \rangle_{\rm SD} &=& \left[1+ \sum_{ma}\rho_{ma} a^\dag_m a_a+ \frac{1}{2} \sum_{mnab} \rho_{mnab}a_m^\dag a_n^\dag a_b a_a\right. \nonumber\\
  &&\left. + \sum_{m \ne v} \rho_{mv} a^\dag_m a_v + \sum_{mna}\rho_{mnva} a_m^\dag a_n^\dag a_a a_v\right] |\Phi_v\rangle,\nonumber \\
&&  
  \label{expansion}
\end{eqnarray}
where $|\Phi_v\rangle$ is the DHF wave function of the state. In the above expression, $a^\dag_i$ and $a_i$ are the creation and 
annihilation operators with the indices $\{m,n\}$ and $\{a,b\}$ designating the virtual and core orbitals of $|\Phi_v\rangle$, 
$\rho_{ma}$ and $\rho_{mv}$ are the corresponding single core and valence excitation coefficients, and $\rho_{mnab}$ and 
$\rho_{mnva}$ are the double core and valence excitation coefficients. To obtain the DHF wave function, we use a finite size basis set 
consisting of 70 B-splines constrained to a large cavity of radius R = $220$ a.u. and solved self consistently using the Roothan equation 
on a nonlinear grid. 

\begin{table*}[t!]\fontsize{7.5}{10.0}\selectfont
\caption{\label{magicd521} The $\lambda_{\rm{magic}}$ values (in nm) with their corresponding polarizabilities $\alpha^J_d(\lambda_{\rm{magic}})$ 
(in a.u.) for the $nS_{1/2}(M_{J}={1/2}) \rightarrow (n-1)D_{5/2}$ transitions using circularly polarized light in the $^{43}$Ca$^+$,$^{87}$Sr$^+$ and $^{137}$Ba$^+$ ions. The resonant wavelengths $\lambda_{\rm{res}}$ to indicate the placement of the $\lambda_{\rm{magic}}$ in between resonant transitions are also listed.}
\setlength{\tabcolsep}{0.1pt}
\begin{tabular}{lccccccccccccccccc}
\hline
\hline
& & \multicolumn{2}{c}{$M_{J}$=5/2}& \multicolumn{2}{c}{$M_{J}$=3/2} & \multicolumn{2}{c}{$M_{J}$=1/2} & \multicolumn{2}{c}{$M_{J}$=$-$1/2} &\multicolumn{2}{c}{$M_{J}$=$-$3/2} & \multicolumn{2}{c}{$M_{J}$=$-$5/2}\\
\hline 
Resonances &$\lambda_{\rm{res}}$ & $\lambda_{\rm{magic}}$ & $\alpha^J_d(\lambda_{\rm{magic}})$ & $\lambda_{\rm{magic}}$ & $\alpha^J_d(\lambda_{\rm{magic}})$& $\lambda_{\rm{magic}}$ &$\alpha^J_d(\lambda_{\rm{magic}})$ &  $\lambda_{\rm{magic}}$ &$\alpha^J_d(\lambda_{\rm{magic}})$ &  $\lambda_{\rm{magic}}$ &$\alpha^J_d(\lambda_{\rm{magic}})$ &  $\lambda_{\rm{magic}}$ &$\alpha^J_d(\lambda_{\rm{magic}})$ \\
\hline \\
$^{43}$Ca$^+$  & \multicolumn{1}{c}{\underline{$4S_{1/2}-3D_{5/2}$}}   \\
$4S_{1/2}-4P_{3/2}$ & 393.366 \\
                    && 394.64(2) & $-25.67$ & 394.64(3) & $-9.69$ & 394.64(2) & 1.0 & 394.63(4) & 6.33 & 394.63(3) & 17.0 & 394.63(3) & 22.39 \\ 
$4S_{1/2}-4P_{1/2}$ & 396.847 & \\
$3D_{5/2}-4P_{3/2}$ & 854.209\\
                    &&  - & - & 1150.39(2) & 85.95 & 975.6(4) & 90.54 & 891.4(3) & 94.09 &-&-&-&-\\
$^{87}$Sr$^+$  & \multicolumn{1}{c}{\underline{$5S_{1/2}-4D_{5/2}$}}   \\
$5S_{1/2}-5P_{3/2}$ & 407.7714 \\
			     &&  412.5(3)  &  $-26.94$ & 412.46(2) & $-10.05$ & 412.42(4) & 6.98 & 412.38(4) & 24.16 & 412.35(3)& 37.15 & 412.31(3)& 54.61  \\
$5S_{1/2}-5P_{1/2}$ & 421.5524 & \\
$4D_{5/2}-5P_{3/2}$ & 1032.7309 & \\ 
                    &&  -& -    & - &  -& 1379.4(3) & 100.44 & 1130.8(4) & 105.46 & - & - & -& -\\
$^{137}$Ba$^+$  & \multicolumn{1}{c}{\underline{$6S_{1/2}-5D_{5/2}$}} \\
$6S_{1/2}-6P_{3/2}$ & 455.4033 \\
			    && 469.1(5) & $-95.43$  & 468.5(1) & $-45.31$ & 467.9(4) & $-3.93$ & 467.6(4) & 29.01 & 467.3(1) & 53.75 & 467.1(1) & 70.27  \\
$6S_{1/2}-6P_{1/2}$ & 493.4077   &\\                    
$5D_{5/2}-6P_{3/2}$ & 614.1713 \\
		    && 810.1(8) & 189.71 & 691(2) & 237.75 & 644.7(5) & 275.99 & 623.3(5) & 302.79 & - & - & -& -  \\
\hline\\
\end{tabular}                    
\end{table*}

\begin{table*}[t!]\fontsize{7.5}{10.0}\selectfont
\caption{\label{magicd522} The $\lambda_{\rm{magic}}$ values (in nm) with their corresponding polarizabilities $\alpha^J_d(\lambda_{\rm{magic}})$ 
(in a.u.) for the $nS_{1/2}(M_{J}=-{1/2}) \rightarrow (n-1)D_{5/2}$ transitions using circularly polarized light in the $^{43}$Ca$^+$,$^{87}$Sr$^+$ and $^{137}$Ba$^+$ ions. The resonant wavelengths $\lambda_{\rm{res}}$ to indicate the placement of the $\lambda_{\rm{magic}}$ in between resonant transitions are also listed.}
\setlength{\tabcolsep}{0.1pt}
\begin{tabular}{lccccccccccccccccc}
\hline
\hline
& & \multicolumn{2}{c}{$M_{J}$=5/2}& \multicolumn{2}{c}{$M_{J}$=3/2} & \multicolumn{2}{c}{$M_{J}$=1/2} & \multicolumn{2}{c}{$M_{J}$=$-$1/2} &\multicolumn{2}{c}{$M_{J}$=$-$3/2} & \multicolumn{2}{c}{$M_{J}$=$-$5/2}\\
\hline 
Resonances &$\lambda_{\rm{res}}$ & $\lambda_{\rm{magic}}$ & $\alpha^J_d(\lambda_{\rm{magic}})$ & $\lambda_{\rm{magic}}$ & $\alpha^J_d(\lambda_{\rm{magic}})$& $\lambda_{\rm{magic}}$ &$\alpha^J_d(\lambda_{\rm{magic}})$ &  $\lambda_{\rm{magic}}$ &$\alpha^J_d(\lambda_{\rm{magic}})$ &  $\lambda_{\rm{magic}}$ &$\alpha^J_d(\lambda_{\rm{magic}})$ &  $\lambda_{\rm{magic}}$ &$\alpha^J_d(\lambda_{\rm{magic}})$ \\
\hline \\
$^{43}$Ca$^+$  & \multicolumn{1}{c}{\underline{$4S_{1/2}-3D_{5/2}$}}   \\
%$4S_{1/2}-4P_{3/2}$ & 393.366 \\
%                    && 394.64(2) & $-25.67$ & 394.64(3) & $-9.69$ & 394.64(2) & 1.0 & 394.63(4) & 6.33 & 394.63(3) & 17.0 & 394.63(3) & 22.39 \\ 
$4S_{1/2}-4P_{1/2}$ & 396.847 & \\
$3D_{5/2}-4P_{3/2}$ & 854.209\\
                    &&  - & - & 1173.5(2) & 85.12 & 982.7(2) & 89.77 & 893.4(3) & 93.37 &-&-&-&-\\
$^{87}$Sr$^+$  & \multicolumn{1}{c}{\underline{$5S_{1/2}-4D_{5/2}$}}   \\
$5S_{1/2}-5P_{3/2}$ & 407.7714 \\
			     &&  416.93(4)  &  $-27.93$ & 416.88(3) & $-9.02$ & 416.85(4) & 7.66 & 416.81(4) & 23.82 & 416.77(1)& 38.63 & 416.74(3)& 52.53  \\
$5S_{1/2}-5P_{1/2}$ & 421.5524 & \\
			    &&  442.49(4)  &  $-31.93$ & 442.36(3) & $-12.11$ & 442.25(3) & 6.12 & 442.13(4) & 22.84 & 442.03(8)& 38.18 & 441.93(3)& 52.07  \\
			    &&  477.99(5)  &   $39.14$ & 478.85(1) & $-16.77$ & 479.32(7) & 3.55 & 479.98(9) & 21.77 & 480.65(1)& 37.83 & 481.28(7)& 51.73  \\
$4D_{5/2}-5P_{3/2}$ & 1032.7309 & \\ 
                    &&  -& -    & - &  -& 1391.4(3) & 98.46 & 1134.5(4) & 102.92 & - & - & -& -\\
$^{137}$Ba$^+$  & \multicolumn{1}{c}{\underline{$6S_{1/2}-5D_{5/2}$}} \\
$6S_{1/2}-6P_{3/2}$ & 455.4033 \\
			    && 479.5(2) & $-108.02$ & 478.9(2) & $-52.89$ & 478.4(2) & $-7.78$  & 477.9(4) & 27.52 & 477.6(4) & 53.33 & 477.4(7) & 69.73  \\
$6S_{1/2}-6P_{1/2}$ & 493.4077   &\\           
			    && 505.5(2) & $-149.78$  & 505.3(4) & $-78.36$ & 505.04(41) & $-20.96$  & 504.8(4) & 25.56 & 504.76(4) & 52.23 & 504.69(7) & 68.53  \\
			    && 568.8(2) & $-451.72$ & 572.14(2) & $-284.62$ & 576.52(2) & $-144.27$ & 582.4(2) & $-31.07$ & 589.7(4) & 50.45 & 591.7(2) & 65.99  \\
$5D_{5/2}-6P_{3/2}$ & 614.1713 \\
		    && 894.3(8) & 153.62 & 729.9(2) & 173.67 & 665.9(3) & 180.26 & 632.6(5) & 171.84 & - & - & -& -  \\
\hline\\
\end{tabular}                    
\end{table*}
In order to verify contributions from the higher level excitations, we also estimate leading order contributions from the triple excitations 
perturbatively in the SD method framework (SDpT method) by expressing the atomic wave functions as \cite{Blundell,theory}
\begin{eqnarray}
|\Psi_v \rangle_{\rm SDpT}=|\Psi_v \rangle_{\rm SD}+\frac{1}{6} \sum_{mnrab} \rho_{mnrvab}^{pert} a_m^\dag a_n^\dag a_r^\dag a_b a_a a_v
|\Phi_v\rangle,\nonumber\\
\label{sdpt}
\end{eqnarray}
where $\rho_{mnrvab}^{pert}$ is the perturbed triple valence excitation amplitudes.

 After obtaining wave functions employing the SD and SDpT method, we determine the E1 matrix element for a given transition between the 
states $|\Psi_v\rangle$ and $|\Psi_w\rangle$ by evaluating the expression
\begin{equation}
 Z_{vw} = \frac{\langle \Psi_v|Z|\Psi_w\rangle}{\sqrt{\langle\Psi_v|\Psi_v\rangle\langle\Psi_w|\Psi_w\rangle}}.
 \label{wavef}
\end{equation}

\begin{table*}[t!]\fontsize{8.0}{10.0}\selectfont
 \caption{\label{magicCa} The $\lambda_{\rm{magic}}$ values (in nm) and their corresponding polarizabilities $\alpha^F_d(\lambda_{\rm{magic}})$ 
 (in a.u.) for the $nS_{1/2}\mid$ ${F}$,$M_{F}\rangle \rightarrow (n-1)D_{{3/2},{5/2}}\mid$ $F'$,$M_{F'}\rangle$ transitions in $^{43}$Ca$^+$ (I=7/2), $^{87}$Sr$^+$ (I=9/2) and 
$^{137}$Ba$^+$ (I=3/2). }
\begin{ruledtabular}
\begin{tabular}{lccccccc}
\multicolumn{6}{c}{{$^{43}$Ca$^+$}}\\
 & \multicolumn{2}{c}{\underline{ $4S_{1/2}-3D_{3/2}$}} &  & \multicolumn{2}{c}{\underline{ $4S_{1/2}-3D_{5/2}$}} \\ 
$\mid$ ${F}$,$M_{F}\rangle \rightarrow \mid$ $F'$,$M_{F'}\rangle$&   $\lambda_{\rm{magic}}$  &  $\alpha^F_d(\lambda_{\rm{magic}})$ & $\mid$ $F$,$M_{F}\rangle \rightarrow \mid$ $F'$,$M_{F'}\rangle$&   $\lambda_{\rm{magic}}$  &  $\alpha^F_d(\lambda_{\rm{magic}})$ \\
$\mid$3,0 $\rangle \rightarrow \mid2,0 \rangle$ & 395.79(2) & 8.69    & $\mid$3,0$\rangle \rightarrow \mid$1,0$\rangle$ &  395.79(3) & 3.35   \\
					         & 852.5(2) & 95.86   &                                               & 992(4)  & 89.71 \\
					         & 999(4) & 89.46     & $\mid$3,0$\rangle \rightarrow \mid$2,0$\rangle$ &  395.79(3) & 3.35 \\
 $\mid$3,0$\rangle \rightarrow \mid$3,0$\rangle$ & 395.79(2) & 3.35   & 						& 1077(9) & 87.31  \\
				                 & 851.3(2)  & 95.95  & $\mid$3,0$\rangle \rightarrow \mid$3,0$\rangle$ &  395.79(3) & 3.35\\ 
				                 & 1089(7)& 87.02     & 						& 1064(9) & 87.61\\
 $\mid$3,0$\rangle \rightarrow \mid$4,0$\rangle$ & 395.79(4) & 3.35   & $\mid$3,0$\rangle \rightarrow \mid$4,0$\rangle$& 395.78(3) & 3.35 \\
					         & 851.7(2) & 95.92   &  						& 1032(7) & 88.48 \\
					         & 1049(4) & 87.99    &$\mid$3,0$\rangle \rightarrow \mid$5,0$\rangle$& 395.79(3) & 3.35 \\
 $\mid$3,0$\rangle \rightarrow \mid$5,0$\rangle$ & 395.79(2) & 8.69   &  						& 992(4) & 89.71\\
					         & 853.4(2) & 95.82   &$\mid$4,0$\rangle$ $\rightarrow$ $\mid$6,0$\rangle$& 395.79(3) & 8.69\\  
					         & 970(3) & 90.44     &                                                   & 949(3) & 91.21\\	 
					         \hline\\
					         \multicolumn{6}{c}{{$^{87}$Sr$^+$}}\\
 & \multicolumn{2}{c}{\underline{ $5S_{1/2}-4D_{3/2}$}}    &  & \multicolumn{2}{c}{\underline{ $5S_{1/2}-4D_{5/2}$}} \\ 
$\mid$F,$M_F\rangle$ $\rightarrow$ $\mid F'$,$M_{F'}\rangle$& $\lambda_{\rm{magic}}$  &  $\alpha^F_d(\lambda_{\rm{magic}})$ & $\mid$F,$M_F\rangle$ $\rightarrow$ $\mid F'$,$M_{F'}\rangle$ &   $\lambda_{\rm{magic}}$  &  $\alpha^F_d(\lambda_{\rm{magic}})$ \\
$\mid$4,0$\rangle$ $\rightarrow$ $\mid$3,0$\rangle$ & 416.99(8) & 14.79  & $\mid$4,0$\rangle$ $\rightarrow$ $\mid$3,0$\rangle$  &  416.99(8) & 13.72\\
			                             & 1017.4(3) & 108.24 & $\mid$4,0$\rangle$ $\rightarrow$ $\mid$4,0$\rangle$ & 416.99(7) & 13.72\\
			                             & 1534(30) & 97.96  & $\mid$4,0$\rangle$ $\rightarrow$ $\mid$5,0$\rangle$  & 416.99(8) & 14.14\\
$\mid$4,0$\rangle$ $\rightarrow$ $\mid$4,0$\rangle$ & 416.99(8) & 13.87 & $\mid$4,0$\rangle$ $\rightarrow$ $\mid$6,0$\rangle$   &  416.99(8) & 14.56\\
			                             & 1010.7(5) & 108.52 &                                                     &  1486(32) & 98.45\\
$\mid$4,0$\rangle$ $\rightarrow$ $\mid$5,0$\rangle$ & 416.99(7) & 14.15 & $\mid$5,0$\rangle$ $\rightarrow$ $\mid$7,0$\rangle$   & 416(3) & 14.56\\
			                            & 1012.4(6) & 108.45 &                                                      & 1301(14) & 100.88\\
$\mid$4,0$\rangle$ $\rightarrow$ $\mid$6,0$\rangle$& 416.99(7) & 14.99\\
			                            & 1020.6(2) & 108.12\\
			                            & 1429(20) & 99.08\\	
			                            \hline\\
\multicolumn{6}{c}{{$^{137}$Ba$^+$}}\\
 & \multicolumn{2}{c}{\underline{ $6S_{1/2}-5D_{3/2}$}} &  & \multicolumn{2}{c}{\underline{ $6S_{1/2}-5D_{5/2}$}} \\
$\mid$ ${F}$,$M_{F}\rangle \rightarrow \mid$ $F'$,$M_{F'}\rangle$&   $\lambda_{\rm{magic}}$  &  $\alpha^F_d(\lambda_{\rm{magic}})$ & $\mid$ $F$,$M_{F}\rangle \rightarrow \mid$ $F'$,$M_{F'}\rangle$&   $\lambda_{\rm{magic}}$  &  $\alpha^F_d(\lambda_{\rm{magic}})$ \\
$\mid$1,0$\rangle \rightarrow \mid$0,0$ \rangle$ & 480.6(7)& $-2.79$  & $\mid$1,0$\rangle \rightarrow \mid$1,0$\rangle$ &   480.53(3) & 5.36\\
			                            & 588.3(2) & 329.33 &    & 641.2(5) & 256.97\\
                                                    & 695.7(9) & 219.39\\
$\mid$1,0$\rangle \rightarrow \mid$1,0$\rangle$ &  480.7(4)& $-9.37$   & $\mid$1,0$\rangle \rightarrow \mid$2,0$\rangle$ &  480.58(4) & 0.91\\
			                            & 587.3(3) & 331.55  & & 647.8(6) & 251.11\\
			                            & 720.47(2) & 207.98\\ 
$\mid$1,0$\rangle \rightarrow \mid$2,0$\rangle$ & 480.68(4) & $-2.88$  & $\mid$1,0$\rangle \rightarrow \mid$3,0$\rangle$ &  480.55(4) & 3.39\\
  			                            & 588.3(5) & 329.33 &  & 644.1(5) & 254.42\\
			                            & 695.7(5) & 219.40\\
$\mid$1,0$\rangle \rightarrow \mid$3,0$\rangle$ & 480.56(4) & 3.55 & $\mid$2,0 $\rangle \rightarrow \mid$4,0$\rangle$  & 480.56(5) & 7.84\\
			                            & 589.7(4) & 329.62 &    & 637.7(3) & 260.25\\
			                             & 675.4(5) & 231.02\\
                                   
\end{tabular}
\end{ruledtabular}		        
\end{table*}

To find out the uncertainties with the calculated E1 matrix elements, we have carried out semi-empirical scaling of the wave functions that 
accounts for evaluation of missing correlations contributions to the wave functions from the approximated SD and SDpT methods. This 
procedure involves scaling of the excitation coefficients and reevaluation of the E1 matrix elements. The scaling factors are decided from 
the correlation energy trends in the SD and SDpT methods. Details regarding this scaling procedure are given in Ref. \cite{safro07}.

\begin{figure}
\includegraphics[width=8.5cm,height=6.0cm]{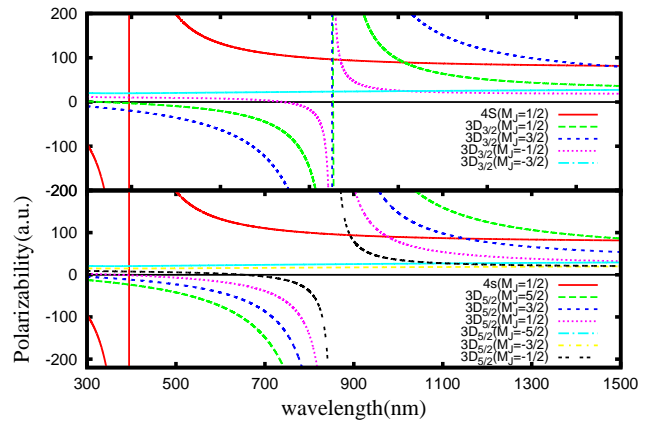}
\caption{(Color online) Dynamic dipole polarizabilities (in a.u.) for the $4S_{1/2}$ and $3D_{{3/2},{5/2}}$ states 
of $^{43}$Ca$^+$ with $A=-1$.}
\label{Casd}     
\end{figure}

\begin{table}
\caption{\label{tuneoutj} The $\lambda_{\rm{T}}$ values (in nm) for all possible magnetic sublevels $M_{J}$ of the $^{43}$Ca$^+$,$^{87}$Sr$^+$ and
$^{137}$Ba$^+$ alkaline earth-metal ions using circularly polarized light.}
\begin{ruledtabular}
\begin{tabular}{lcc|lcc|lcc}
\multicolumn{3}{c}{$^{43}$Ca$^+$} & \multicolumn{3}{c}{$^{87}$Sr$^+$}& \multicolumn{3}{c}{$^{137}$Ba$^+$}  \\
State& $M_{J}$ & $\lambda_{\rm{T}}$ & State& $M_{J}$ & $\lambda_{\rm{T}}$  & State& $M_{J}$ & $\lambda_{\rm{T}}$\\
\hline
%\cline{1-3}\cline{3-3} \cline{4-6}\cline{6-6}\cline{7-9}\\
$4S_{1/2}$&$1/2$ & 482.9   & $5S_{1/2}$& $-1/2$ & 479.2  &$6S_{1/2}$& $-1/2$ & 584.7\\
          &       &        &           & $-1/2$ & 416.8 &         & $-1/2$ & 478.3\\
          &       &        &           & $-1/2$ & 442.3 &         & $-1/2$ & 504.9\\
	  &       &        &           & $1/2$ & 417.2  &         &$1/2$ & 482.9\\
$3D_{3/2}$&$-1/2$ & 746.2   & $4D_{3/2}$&$-1/2$ & 881.5   & $5D_{3/2}$&$-1/2$ &  551.6\\
           &$1/2$ &  853.9 &          &$1/2$ & 1024.7  &          &$1/2$ & 424.4\\
	  &$3/2$ & 851.2   &           & $1/2$ & 467.6 &          &$3/2$ & 590.0\\
                    &&&                &$3/2$ & 1010.1  &\\
$3D_{5/2}$&$-1/2$ & 690.2 & $4D_{5/2}$  &$-1/2$ & 834.5 &$5D_{5/2}$&$-1/2$ & 571.5\\
          & $1/2$ & 434.5 &              & $1/2$ & 525.6&       &$1/2$ & 489.3\\
          &&&                   &&&                              &$3/2$ & 417.5\\		     
\end{tabular}
\end{ruledtabular}		        
\end{table}

\section{Results and Discussion} 

\begin{table}
\caption{\label{tuneoutf} The $\lambda_{\rm{T}}$ values (in nm) for $F$  hyperfine levels ($M_{F}$=0) of the $^{43}$Ca$^+$,$^{87}$Sr$^+$ 
and $^{137}$Ba$^+$ alkaline earth-metal ions using circularly polarized light.}
\begin{ruledtabular}
\begin{tabular}{lcc|lcc|lcc}
\multicolumn{3}{c}{$^{43}$Ca$^+$} & \multicolumn{3}{c}{$^{87}$Sr$^+$}  &\multicolumn{3}{c}{$^{137}$Ba$^+$} \\ 
%$^{43}$Ca$^+$ & &&$^{87}$Sr$^+$  &&& $^{137}$Ba$^+$\\
State& $F$ & $\lambda_{\rm{T}}$ & State& $F$ & $\lambda_{\rm{T}}$ &State& $F$ & $\lambda_{\rm{T}}$\\
\hline
%\multicolumn{3}{c}{$^{43}$Ca$^+$} & \multicolumn{3}{c}{$^{87}$Sr$^+$}  &\multicolumn{3}{c}{$^{137}$Ba$^+$} \\  \\
%$^{43}$Ca$^+$ & &&$^{87}$Sr$^+$  &&& $^{137}$Ba$^+$\\
%\cline{1-3}\cline{3-3} \cline{4-6}\cline{6-6}\cline{7-9}\\
 $4S_{1/2}$ &$3$ & 482.9   &$5S_{1/2}$ &$4$ & 417.1  &$6S_{1/2}$&$1$ & 480.5 \\
 $3D_{3/2}$&$2$ & 505.7    &$4D_{3/2}$&$3$ &616.3     &$5D_{3/2}$ &$0$ &472.3\\ 
                &&853.4                && &1023.2                &   && 597.9\\
            &$3$ &467.3                &&$4$ & 570.7             &&$1$ & 456.6\\
             &  &881.6&                &&1012.9		    &   && 591.9\\
            &$4$ & 480.8               &&$5$ & 583.2             &&$2$ & 472.4\\
                & &852.2               &  & & 1015.5             &   && 597.9\\
            &$5$ & 526.4              &&$6$ & 634.8		 &&$3$ & 491.3\\
               & & 854.5               &  & &1028.5		 &   && 608.7\\
 $3D_{5/2}$&$1$ & 495.8  & $4D_{5/2}$&$3$ & 560.4    & $5D_{5/2}$ &$1$ & 552.5\\
&$2$ & 456.1			&&$4$ & 557.9                     &&$2$ & 509.5\\
&$3$ & 460.2			&&$5$ & 570.9			  &&$3$ & 519.2\\
&$4$ & 473.5 			&&$6$ & 596.8			  &&$4$ & 527.1\\	
&$5$ &  495.8			&&$7$ & 641.9 & & & \\
&$6$ & 532.5 & & & & & &\\  
\end{tabular}
\end{ruledtabular}		        
\end{table}

Since valence correlation contributions are vital for accurate estimate of polarizabilities, we include the E1 matrix elements among 
the low-lying states up to $4S-7P$, $3D - 7P$ and $3D-6F$ transitions in $^{43}$Ca$^+$, $5S-8P$, $4D-8P$ and $4D-6F$ transitions in $^{87}$Sr$^+$
and $6S-8P$, $5D-8P$ and $5D-6F$ transitions in $^{137}$Ba$^+$ for the evaluation of the ``Main'' contributions. All these matrix elements
are calculated using the RCC method described in the previous section. A few E1 matrix elements of the $nD-nF$ transitions of 
$^{137}$Ba$^+$ are used from Ref. \cite{SahooBa}. We present these considered E1 matrix elements in the Supplemental Material from different levels 
of approximation in the considered many-body methods. Scaled values to the SD and SDpT results, which mainly ameliorates the results by 
accounting corrections beyond the Brueckner-orbital contributions \cite{Blundell}, are also given in the same table. We then give the 
``Final'' results considering the most reliable values along with their estimated uncertainties in parentheses in the last column of this table~\cite{safroca}. 
As can be seen from the Supplemental Material, the DHF method gives large E1 matrix element values while the SD method brings down the values while
the SDpT method slightly increases the values from the SD values. The scaled values from both the SD and SDpT methods modifies these values 
marginally. Thus, these values seem to be very reliable for determining the polarizabilities of the considered ions. For the final use,
we recommend the SD method values and uncertainties to these values are estimated by taking the differences from the results obtained 
using the SDpT method. Below, we discuss polarizability results using these E1 matrix elements and magic wavelengths of the $S-D$ clock 
transitions of the above alkaline-earth ions.

\subsection{Static Polarizability results} 

Using the E1 matrix elements given in Supplement Material, we first evaluate the static polarizabilities of the ground and $(n-1)D_{J}$ states of 
the $^{43}$Ca$^+$, $^{87}$Sr$^+$ and $^{137}$Ba$^+$ ions and compare them with the previously available experimental and theoretical results in
Table \ref{pol1}. We give both the scalar and tensor polarizabilities of the considered ground and $(n-1)D_{J}$ states along with their 
contributions from ``Main'' and ``Tail'' to valence  $\alpha_{d,i}^{J,v}$, core-valence  $\alpha_{d,i}^{J,cv}$ and core $\alpha_{d,i}^{J,c}$
contributions to our calculations in this table. These results are discussed ion-wise below.

\textbf{$^{43}$Ca$^+$}: As can be seen from Table \ref{pol1}, $\alpha_{d,i}^J$(0) value of 76.1(2) a.u. for the ground state polarizability 
obtained for $^{43}$Ca$^+$ ion is in close agreement with the other theoretical calculations which are 75.28 a.u. and 75.49 a.u. by Tang 
\textit{et al.} \cite{Tang} and Mitroy \textit{et al.} \cite{mitroyca}, respectively. Mitroy \textit{et al.} have also given these values in 
the $3D_{3/2}$ and $3D_{5/2}$ states of Ca$^+$. They had evaluated these polarizabilities by diagonalizing a semi-empirical Hamiltonian 
constructed in a large dimension single electron basis. Our estimated values agree quite well those values within the quoted error bars. In
Ref. \cite{sahooca09}, {\it ab initio} calculations of these quantities are reported using a RCC method and our values are also compare 
reasonably well with them. The most stringent experimental value for the $^{43}$Ca$^+$ ground state polarizability was obtained by spectral
analysis in Ref. \cite{edward}, our result is also in agreement with this value.

\textbf{$^{87}$Sr$^+$}: Next, we compare our polarizability results for the $^{87}$Sr$^+$ ion given in Table \ref{pol1}. The RCC results of the 
$S$ and $D$ states reported by Sahoo \textit{et al.} \cite{sahoo} are in agreement with our values. Mitroy \textit{et al.} have also given these 
values by employing a non-relativistic method in the sum-over-states approach \cite{SR}. It can be seen from Table \ref{pol1} that our ground 
state dipole polarizability is in very good agreement with their result. However, it seems inappropriate to compare their non-relativistic values 
for the dipole polarizabilities of the ${4D}_{J}$ states with our relativistic calculations. The estimate for the ground state static 
polarizability of $^{87}$Sr$^+$ by Barklem \textit{et al.} \cite{SR} derived by combining their theoretical calculations with the experimental 
data from Ref. \cite{Barklem}. There is a considerable discrepancy between their result with our present value. This is mainly because
of omission of the core contribution which has been included by us using the RPA method. There are no direct experimental results available for
the Sr$^+$ ion dipole polarizabilities to make a comparative analysis with the theoretical values.

\textbf{$^{137}$Ba$^+$}: Our precise ground state polarizability calculation gives a value of 123.2(5) a.u. for the $^{137}$Ba$^+$ ion, which 
is in good agreement with the high precision measurement achieved by a novel technique based on the resonant excitation Stark ionization 
spectroscopy \cite{snow}. We also expect that results of the ${5D}_{J}$ states will also be of similar accuracies in this ion.

Having analyzed accuracies of the static polarizabilities satisfactorily, we now move on to present the dynamic polarizabilities in the above 
ions. We adopt the similar procedures for calculations of these quantities, thus anticipating similar accuracies in the dynamic polarizability 
values as their corresponding static values. This ascertains us to determine the $\lambda_{\rm{magic}}$ values of the $nS-(n-1)D_{3/2}$, 
$nS-(n-1)D_{5/2}$ transitions and the $\lambda_{\rm T}$ values of the associated states in the alkaline earth metal ions without
much qualm using these dynamic polarizabilities.

\subsection{Dynamic dipole polarizabilities at 1064 nm}
 
 We are discussing here on the dynamic polarizabilities at the 1064 nm that are very demanding for creating high-field seeking traps of the 
considered ions (far detuned traps), where the atoms are attracted to the intensity maxima. Recently Chen \textit{et al.} \cite{Chen} had carried 
out measurements of scalar and tensor contributions to the atomic polarizabilities in the Rb atom at this wavelength. In our calculation 
of dynamic polarizabilities in the considered $^{43}$Ca$^+$, $^{87}$Sr$^+$ and $^{137}$Ba$^+$ ions, we have used same E1 matrix elements to 
determine the ``Main'' contributions and estimated other non-dominant contributions in the similar procedure as was done for the evaluation of 
the static polarizabilities. We list the contributions to the $nS_{1/2}$ and $(n-1)D_{J}$ dynamic polarizabilities at this wavelength of the above ions in Table \ref{pol2}. 
The dominant contributions are listed explicitly. This table illustrates very fast convergence of the $nS_{1/2}$ state polarizabilities from 
which we find that the largest contributions are appearing from the $nP$ excited states. We also notice that contributions from the 
$4D_{3/2}$-$5P_{1/2}$ transition to the $4D_J$ state polarizabilities in $^{87}$Sr$^+$ increase 20 times as compared to the contribution from 
this transition to the $4D_J$ state static polarizability. The reason for the overwhelming contribution from this particular transition is
due to the proximity of the $4D_{3/2}$-$5P_{1/2}$ resonance that lies at 1091 nm to the laser wavelength of 1064 nm. 

\subsection{Magic and Tune out wavelengths for circularly polarized light}

\begin{figure}[h]
\includegraphics[width=8.5cm,height=6.0cm]{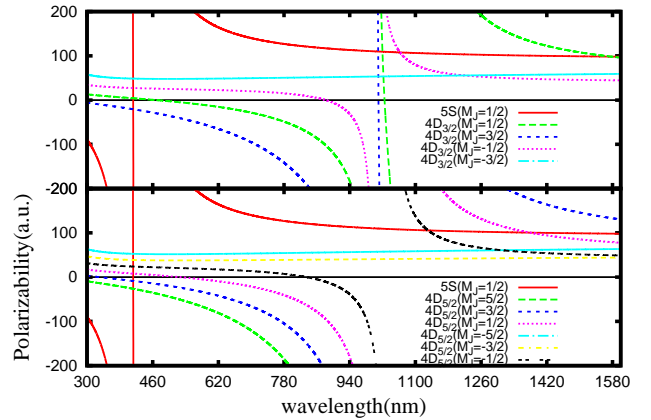}
\caption{(Color online) Dynamic dipole polarizabilities (in a.u.) for the $5S_{1/2}$ and $4D_{{3/2},{5/2}}$ states 
of $^{87}$Sr$^+$ with $A =-1$.}
\label{Srsd}     
\end{figure}

\begin{figure}[h]
\includegraphics[width=8.5cm,height=6.0cm]{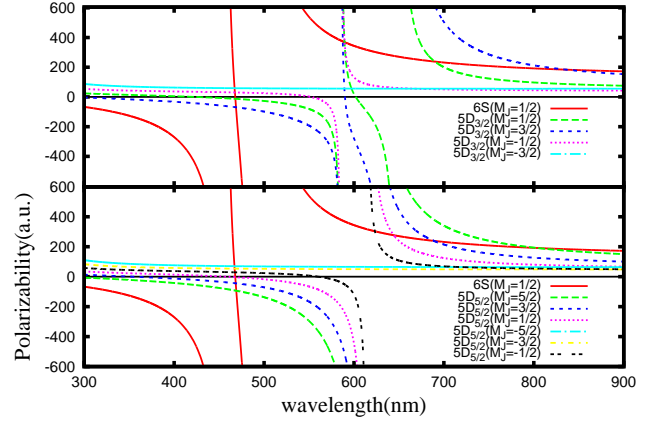}
\caption{(Color online) Dynamic dipole polarizabilities (in a.u.) for the $6S_{1/2}$ and $5D_{{3/2},{5/2}}$ states 
of $^{137}$Ba$^+$ with $A=-1$.}
\label{Basd}     
\end{figure}

In order to find out $\lambda_{\rm{magic}}$ among the $(J,M_{J})$ levels of the $nS_{1/2} \rightarrow (n-1)D_{3/2}$ and $nS_{1/2} 
\rightarrow (n-1)D_{5/2}$ transitions, we plot total dynamic dipole polarizabilities for the $nS_{1/2}$ and $(n-1)D_{{3/2},{5/2}}$ states
in Figs. (\ref{Casd}), (\ref{Srsd}) and (\ref{Basd}) for the $^{43}$Ca$^+$, $^{87}$Sr$^+$ and $^{137}$Ba$^+$ ions respectively. The
$\lambda_{\rm{magic}}$ for the clock transitions are obtained by locating the crossing points between the two polarizability curves. In 
Tables \ref{magicd321}, \ref{magicd322}, \ref{magicd521} and \ref{magicd522}, we list the $\lambda_{\rm{magic}}$ for the considered transitions along with their respective 
uncertainties in the parentheses. The corresponding polarizability values at $\lambda_{\rm{magic}}$ are listed as well. The resonant
wavelengths $\lambda_{\rm{res}}$ are listed in the same table to demonstrate placement of a $\lambda_{\rm{magic}}$ in between 
two resonant transitions. In this work, we use left-handed circularly polarized light ($A$=$-$1) for all purposes considering all possible
positive and negative $M_{J}$ sublevels for the ground $S_{1/2}$ and $D_{3/2,5/2}$ states. 
%[{\bf This has to be given for both the $M_J$ because of the following statement}]. 
Note that $\lambda_{\rm{magic}}$ for the right circularly polarized light of a transition for a given $M_{J}$ are equal to the $\lambda_{\rm{magic}}$ for 
left circularly polarized light with opposite sign of $M_{J}$~\cite{sukhjitJPB}.

\begin{figure}
\includegraphics[width=8.5cm,height=6.0cm]{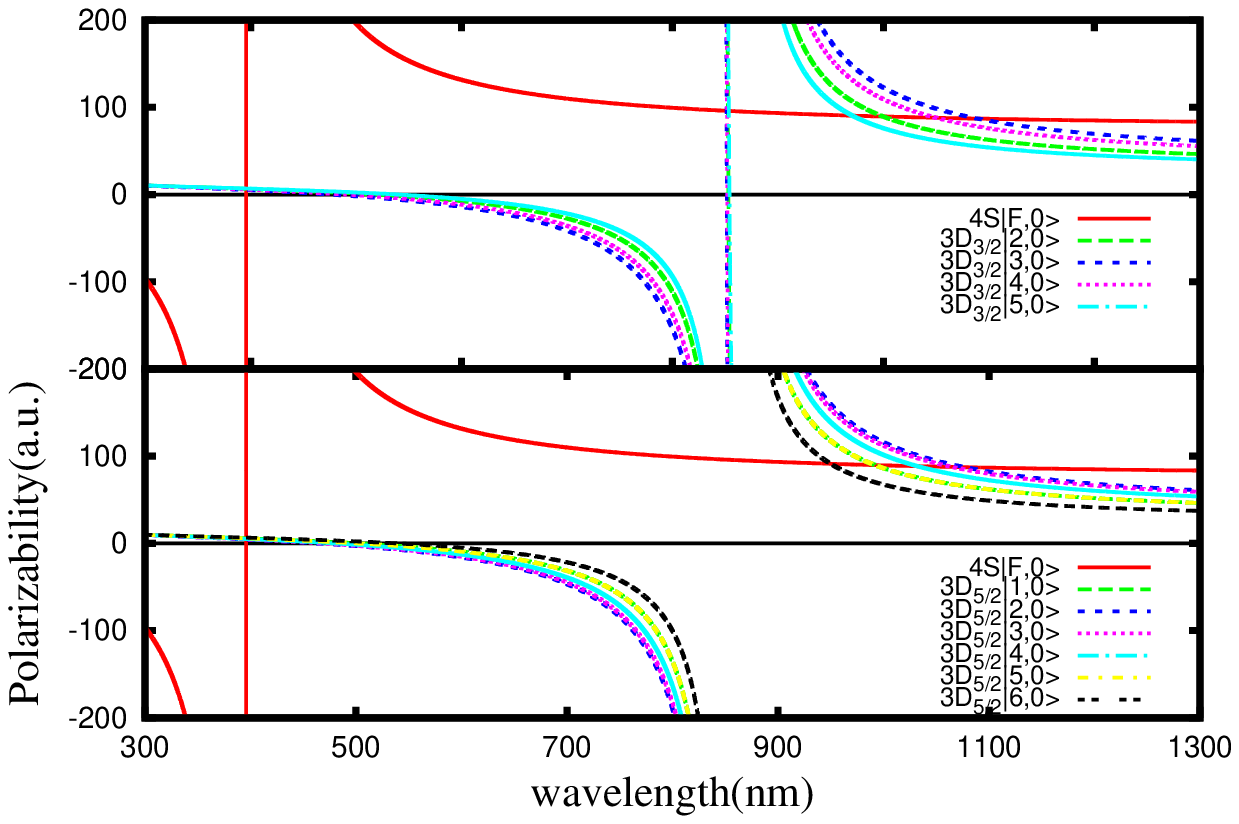}
\caption{(Color online) ${F}$ dependent dynamic dipole polarizabilities (in a.u.) for the $ \mid 4S_{1/2} F M_F=0\rangle$ ($F = $ 3, 4)
and $ \mid 3D_{{3/2},{5/2}}$ ${F} M_F=0\rangle$ states of $^{43}$Ca$^+$ with $A =-1$.} 
\label{Ca1} 
\end{figure}

\begin{figure}[h]
\includegraphics[width=8.5cm,height=6.0cm]{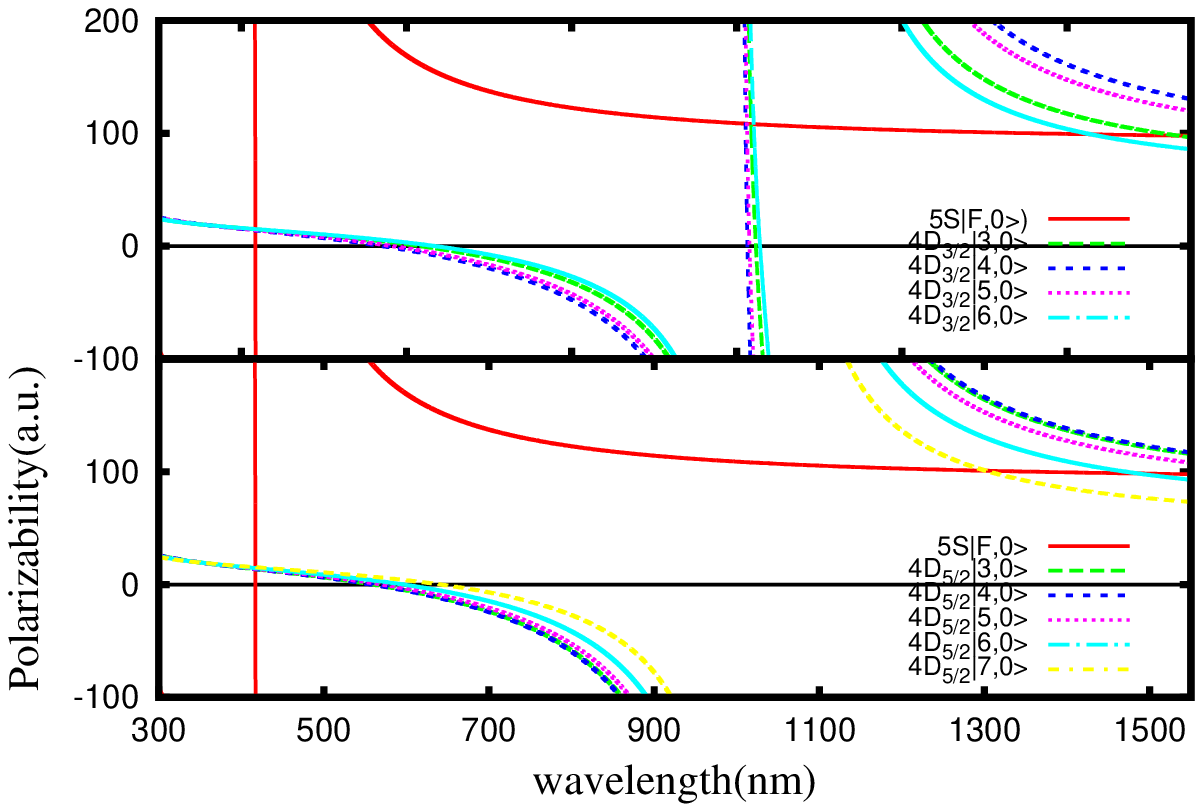}
\caption{(Color online) ${F}$ dependent dynamic dipole polarizabilities (in a.u.) for the $  \mid 5S_{1/2} {F},M_F=0\rangle$ ($F$= 4, 5) and 
$ \mid 4D_{{3/2},{5/2}}$ ${F},M_F=0\rangle$ states of $^{87}$Sr$^+$ with $A =-1$.} 
\label{Sr1}     
\end{figure}

\begin{figure}[h]
\includegraphics[width=8.5cm,height=6.0cm]{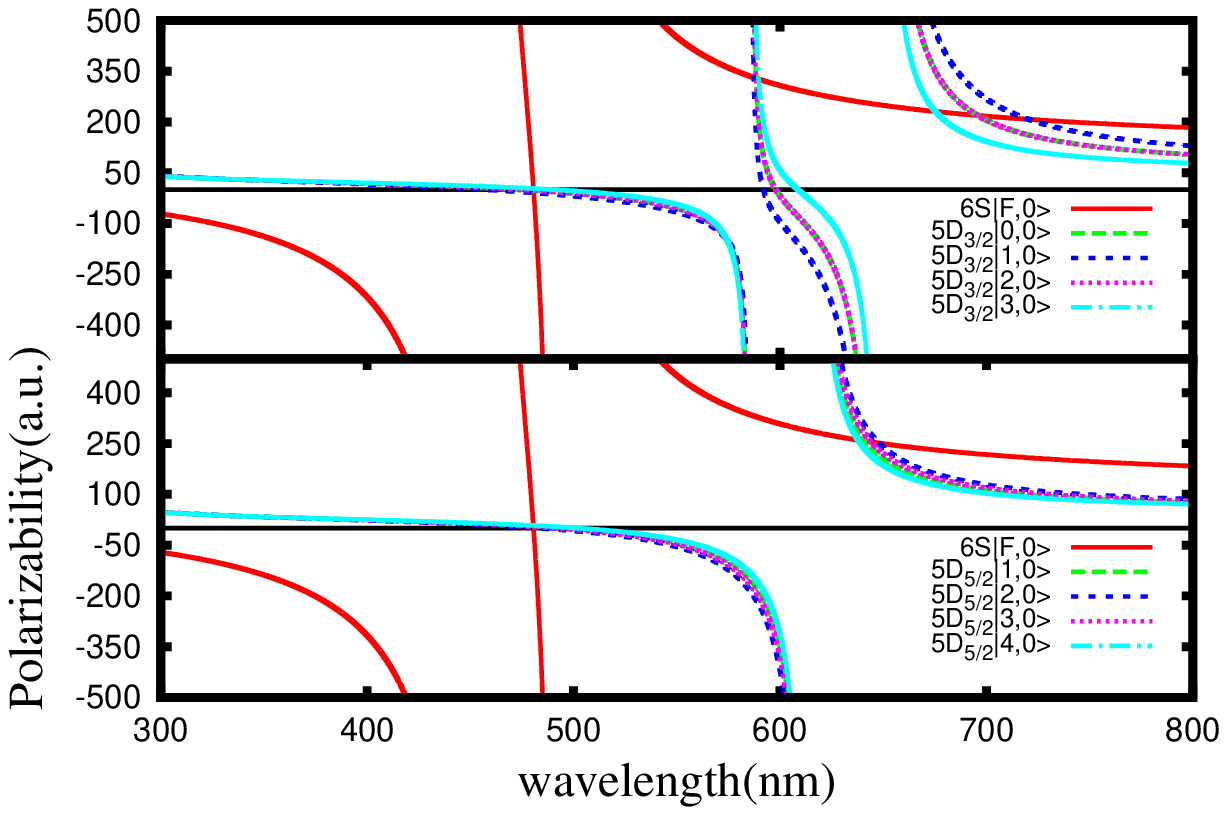}
\caption{(Color online)  ${F}$ dependent dynamic dipole polarizabilities (in a.u.) for the $ \mid6S_{1/2} F,M_F0\rangle$ ($F$ = 1, 2) and 
$\mid5D_{{3/2},{5/2}}$ ${F},M_F=0\rangle$ states of $^{137}$Ba$^+$ with $A =-1$.} 
\label{Ba1}     
\end{figure}

We also investigate $\lambda_{\rm{magic}}$ between the transitions involving the $\mid{n}S_{1/2} {F},M_{F}=0\rangle$ and $\mid(n-1)D_{{3/2},{5/2}}{F},M_{F}=\rm{0}\rangle$ 
states. The ac Stark shifts of the hyperfine levels of an atomic state are calculated using the method described in Sec. \ref{thesec}. 
We choose $M_F=0$ sublevels in the hyperfine transitions as for this particular magnetic sublevel, the first order Zeeman shift vanishes. 
This is advantageous for the optical clock experiments \cite{Champenois}. We show the $\lambda_{\rm{magic}}$ values of the 
$\mid nS_{1/2}{F},M_{F}=0\rangle \rightarrow \mid(n-1)D_{{3/2},{5/2}}{F},M_{F}=\rm{0}\rangle$ transitions in Figs. (\ref{Ca1}), 
(\ref{Sr1}) and (\ref{Ba1}) for the $^{43}$Ca$^+$, $^{87}$Sr$^+$ and $^{137}$Ba$^+$ ions respectively. These values are listed in 
Table~\ref{magicCa} and we discuss below about these results for the individual ion.  

\begin{table*}[t!]
\begin{ruledtabular}
\caption{\label{magicsd} The $\lambda_{\rm{magic}}$ values (in nm) independent of $M_{J}$, $F$ and $M_F$ with their corresponding 
polarizabilities (in a.u.) for the $nS_{1/2}$ - $(n - 1)D_{{3/2},{5/2}}$ transition in the $^{43}$Ca$^+$ ,$^{87}$Sr$^+$ and $^{137}$Ba$^+$ ions 
with $n$=4,5,6 respectively.} 
\begin{tabular}{lccccccc}
\multicolumn{2}{c}{Ca$^+$}  &\multicolumn{2}{c}{Sr$^+$}  & \multicolumn{2}{c}{Ba$^+$}\\
\multicolumn{2}{c}{\underline{ $4S_{1/2}-3D_{3/2}$}}   &\multicolumn{2}{c}{\underline{ $5S_{1/2}-4D_{3/2}$}} & \multicolumn{2}{c}{\underline{ $6S_{1/2}-5D_{3/2}$}} \\ 
 $\lambda_{\rm{magic}}$ &$\alpha^J_{d,0}(\lambda_{\rm{magic}})$ &  $\lambda_{\rm{magic}}$ &$\alpha^J_{d,0}(\lambda_{\rm{magic}})$ &  $\lambda_{\rm{magic}}$ &$\alpha^J_{d,0}(\lambda_{\rm{magic}})$ \\
 
                           395.82(3)   & 4.90     & 416.9(3)    & 14.47  & 480.6(5)   & $-$2.89\\   
			   852.45(2)   & 95.87    & 1014.6(2)   & 108.35 &  588.4(3)  & 329.33  \\
                           1029.7(2)   & 88.55    & -           & -      &   655.50(3)  & 244.89 \\
\multicolumn{2}{c}{\underline{ $4S_{1/2}-3D_{5/2}$}}  & \multicolumn{2}{c}{\underline{ $5S_{1/2}-4D_{5/2}$}} & \multicolumn{2}{c}{\underline{ $6S_{1/2}-5D_{5/2}$}} \\
                           395.82(2)   & 4.20   &   416.9(3)  & 13.3  &  480.6(4) & $-$3.75 \\
                           1014.10(3)  & 89.01  &  1566.2(3) & 97.63  &  695.7(3) & 219.4\\
\end{tabular}
\end{ruledtabular}
\end{table*}

 \begin{table}[t!]
\caption{\label{tunesd}The $\lambda_{\rm{T}}$ values (in nm) independent of $M_{J}$, $F$ and $M_F$ for the $nS_{1/2}$, $(n - 1)D_{3/2}$ and $(n-1)D_{5/2}$ 
states of the $^{43}$Ca$^+$ ,$^{87}$Sr$^+$ and $^{137}$Ba$^+$ ions with $n$=4,5,6 respectively.} 
\begin{ruledtabular}
\begin{tabular}{lc|lc|lc}
%&\multicolumn{6}{c}{$\lambda_{\rm{T}}$(in nm)}\\
\multicolumn{2}{c}{$^{43}$Ca$^+$} &\multicolumn{2}{c}{$^{87}$Sr$^+$}  & \multicolumn{2}{c}{$^{137}$Ba$^+$}\\
State & $\lambda_{\rm{T}}$ & State &  $\lambda_{\rm{T}}$ & State &  $\lambda_{\rm{T}}$\\
\hline
 $4S_{1/2}$ &  395.796  &$5S_{1/2}$  & 417.025  &$6S_{1/2}$&480.596 \\ 
 $3D_{3/2}$& 492.572     &$4D_{3/2}$ &598.633  &$5D_{3/2}$ &472.461\\ 
               &852.776 &&1018.873   && 597.983\\

 $3D_{5/2}$ & 482.642  &$4D_{5/2}$ & 585.677 &$5D_{5/2}$ & 509.687\\
     
\end{tabular}
\end{ruledtabular}		        
\end{table}

{\bf $^{43}$Ca$^+$:} As evident from Fig. (\ref{Casd}), the dynamic polarizabilities for the $4S_{1/2}$ state are small except in the
vicinity of the resonant $4S_{1/2}$-$4P_{1/2}$ and $4S_{1/2}$-$4P_{3/2}$ transitions around 396.847 nm and 393.366 nm respectively. Since 
the $3D_{{3/2},{5/2}}$ states have significant contributions from the resonances in the interested wavelength range, the 
$\lambda_{\rm{magic}}$ are expected to lie in between these resonances. We found a total of nine $\lambda_{\rm{magic}}$ for all possible magnetic 
sublevels of the $4S_{1/2}$-$3D_{3/2}$ transition in between four resonances. The first four $\lambda_{\rm{magic}}$ are located around 395 nm 
between the resonant $4S_{1/2}$-$4P_{1/2}$ and $4S_{1/2}$-$4P_{3/2}$ transitions for all the $M_{J}$ magnetic sublevels of the $4D_{3/2}$ state.
Out of these, two $\lambda_{\rm{magic}}$ support blue-detuned trap, whereas the other two support red-detuned trap. The next five 
$\lambda_{\rm{magic}}$ are identified at 850.9(2) nm, 853.1(2) nm, 1467.8(4) nm, 1013.4(5) nm and 870.7(3) nm, which lie in the infrared region. 
For some of the $M_{J}$ sublevels, the $\lambda_{\rm{magic}}$ is missing. In such case, it would be imperative to consider a geometry where 
$\lambda_{\rm{magic}}$ will be independent of the magnetic sublevels as mention in Sec. \ref{introsec} and discussed elaborately in our previous
work \cite{ab1,ab2}. In Table \ref{magicd521}, we present the $\lambda_{\rm{magic}}$ for the $4S_{1/2}-3D_{5/2}$ transition 
considering all the magnetic sublevels of the $3D_{5/2}$ state. As seen, most of these $\lambda_{\rm{magic}}$ in the $^{43}$Ca$^+$ ion 
support red detuned trap, indicated by small positive values of the polarizabilities for the corresponding $\lambda_{\rm{magic}}$ values. 
Similarly, we tabulate $\lambda_{\rm{magic}}$ for the $4S$ ($M_J$ =$-1/2$) - $3D_{3/2,5/2}$ transition in table \ref{magicd322} and \ref{magicd522}. 
It can be evidently seen from the table that $\lambda_{\rm{magic}}$ are red shifted from the $\lambda_{\rm{magic}}$ for $4S$ ($M_{J} =1/2$) - $3D_{3/2,5/2}$.

Similarly in Table \ref{magicCa}, we list the $\lambda_{\rm{magic}}$ values between 300-1300 nm for the $ \mid4S_{1/2}F,M_{F}=0\rangle \rightarrow 
\mid3D_{{3/2},{5/2}}F,M_{F}=0\rangle$ transitions. The $F$-dependent polarizabilities values at the respective $\lambda_{\rm{magic}}$ are 
listed as well. For this wavelength range, total twenty four $\lambda_{\rm{magic}}$ are located. Out of which, fourteen $\lambda_{\rm{magic}}$ 
are identified in the infrared region. From this table, it can be found that for the $\mid 4S_{1/2} {F},M_{F} = 0\rangle \rightarrow 
\mid3D_{{3/2},_{5/2}}{F},M_{F} = 0\rangle$ transitions, all $\lambda_{\rm{magic}}$ support red-detuned traps.

{\bf $^{87}$Sr$^+$:} The dynamic polarizabilities for the $5S_{1/2}$, $4D_{3/2}$ and $4D_{5/2}$ states of $^{87}$Sr$^+$ calculated by us 
are plotted in Fig. (\ref{Srsd}). A number of $\lambda_{\rm{magic}}$ are identified by the intersections of the polarizability curves of the 
$5S_{1/2}$ and $4D_{{3/2},{5/2}}$ states for all their magnetic sublevels of the $4D_{{3/2},{5/2}}$ states in the $5S_{1/2}(M_{J}=1/2)-4D_{{3/2},{5/2}}$ 
transitions and are presented in Table \ref{magicd321} along with their resonant lines. Four $\lambda_{\rm{magic}}$ are found to be around 
413 nm between the $5S_{1/2}$-$5P_{1/2}$ and $5S_{1/2}$-$5P_{3/2}$, resonant transitions. These values belong to the visible region, while the 
other five $\lambda_{\rm{magic}}$ are located at 1009.3(3) nm, 1019.7(3) nm, 1062.5(3) nm and 1577.2(3) nm that lie in the infrared region. All the
$\lambda_{\rm{magic}}$ values mentioned for the $5S_{1/2}$-$4D_{3/2}$ transition, except the one at 412.5(9) nm, support the red-detuned 
trapping scheme. Similarly, for the $5S_{1/2}$-$4D_{5/2}$ transition total eight $\lambda_{\rm{magic}}$ are appearing for all possible 
$M_{J}$ sublevels. Among them two $\lambda_{\rm{magic}}$ are located at 1379.4(3) nm and 1130.8(4) nm appear after the $4D_{5/2}$-$5P_{3/2}$ 
resonance for the $M_{J}$=1/2 and $M_{J}$=$-1/2$ magnetic sublevels respectively. Use of these $\lambda_{\rm{magic}}$ are recommended to
carry out experiments selectively for the corresponding magnetic sublevels.
Similarly, we present $\lambda_{\rm{magic}}$ for the $5S$ ($M_J$ =$-1/2$) - $4D_{3/2,5/2}$ transition in table \ref{magicd322} and \ref{magicd522}. 
It can be noticed from these tables that $\lambda_{\rm{magic}}$ are red shifted from the $\lambda_{\rm{magic}}$ for 
$5S$ ($M_{J} =1/2$) - $4D_{3/2,5/2}$ transition. We have also determined total sixteen extra $\lambda_{\rm{magic}}$ between $5S_{1/2}$-$5P_{1/2}$ 
and $4D_{3/2}$-$5P_{3/2}$ resonance transition.
      
      In Table~\ref{magicCa}, $\lambda_{\rm{magic}}$ values above 300 nm are listed in the case of $^{87}$Sr$^+$ for the $\mid5S_{1/2}F,M_{F}=
0\rangle \rightarrow \mid4D_{{3/2},{5/2}}F,M_{F}=0\rangle$ transitions. It is found that $\lambda_{\rm{magic}}$ around 417 nm with very small 
polarizabilities for the $\mid5S_{1/2} F M_{F}=0\rangle \rightarrow \mid4D_{3/2}F,M_{F}=0\rangle$ 
and $\mid5S_{1/2} {F} {M_{F}}=0\rangle \rightarrow \mid4D_{5/2} F,M_{F}=0\rangle$ transitions. Therefore, it will be challenging to trap 
the $^{87}$Sr$^+$ ion at these wavelengths. However, the $\lambda_{\rm{magic}}$ values in the infrared region for these transitions may be 
useful for trapping the above ion in the experiments.

{\bf $^{137}$Ba$^+$:} Total nine $\lambda_{\rm{magic}}$ are found for the $6S_{1/2}$-$5D_{3/2}$ transition of $^{137}$Ba$^+$, among which four 
$\lambda_{\rm{magic}}$ are around 468 nm in the vicinity of the $6S_{1/2}$-$6P_{3/2}$ resonant transition. The next $\lambda_{\rm{magic}}$ at
587.6(9) nm, 589.5(3) nm and 589.6(5) nm are located at the sharp intersection of polarizability curves close to the $5D_{3/2}$-$6P_{3/2}$ and 
$5D_{3/2}$-$6P_{1/2}$ resonances, as seen in Fig. (\ref{Ba1}). The last two $\lambda_{\rm{magic}}$ are located at 841.7(5) nm and 690.7(7) nm 
for the $M_{J}$=$3/2$ and $1/2$ sublevel respectively, have positive polarizabilities. Hence these $\lambda_{\rm{magic}}$ could provide 
sufficient trap depth at the reasonable laser power. In fact, some of the expected $\lambda_{\rm{magic}}$ are missing for the $M_{J}$=
$-3/2$ and $-1/2$ sublevels. Similarly, several $\lambda_{\rm{magic}}$ are also located for the $6S_{1/2}$-$5D_{5/2}$ transitions, as seen 
from Fig. (\ref{Ba1}), in the wavelength range 300-800 nm which are listed in Table \ref{magicd521}. The expected trend of locating 
$\lambda_{\rm{magic}}$ between the resonances in this transition is similar to the previous two ions. 
For the $6S$ ($M_J = -1/2$) - $5D_{3/2,5/2}$ transition, we list the $\lambda_{\rm{magic}}$ in table \ref{magicd322} and \ref{magicd522}. 
These magic wavelengths are slightly red shifted to those demonstrated for $6S$ ($M_J = -1/2$) - $5D_{3/2,5/2}$ transition.
We also found total fifteen $\lambda_{\rm{magic}}$ between $6S_{1/2}$-$6P_{1/2}$ and $5D_{3/2}$-$6P_{1/2}$ resonance transitions in both the tables.
We have also determined an extra $\lambda_{\rm{magic}}$ at 590.5 nm in table \ref{magicd322}, which supports a red detuned trap.

    In Table \ref{magicCa}, we also list $\lambda_{\rm{magic}}$ for the $\mid6S_{1/2}{F},{M_{F}}=0\rangle \rightarrow  \mid5D_{{3/2},{5/2}}{F},{M_{F}}=0\rangle$ transitions which 
lie within the wavelength range of 300-800 nm. We correspondingly locate twenty $\lambda_{\rm{magic}}$ in the visible region. We notice that
all the $\lambda_{\rm{magic}}$ values, except around 480 nm, are expected to be more promising for experiments. An ion trap at these  
wavelengths can have sufficient trap depth at the reasonable laser power. 

{\bf Tune-out wavelengths:} Table \ref{tuneoutj} and \ref{tuneoutf} illustrate the identified tune out wavelengths of the $nS_{1/2}$, 
$(n-1)D_{3/2}$ and $(n-1)D_{5/2}$ states in the $(J,M_{J})$ and $(F,M_{F})$ levels of the $^{43}$Ca$^+$, $^{87}$Sr$^+$ and $^{137}$Ba$^+$ alkaline earth metal 
ions. To locate these tune out wavelengths, we have calculated the dynamic polarizabilities of the above states for a particular range of 
wavelength in the vicinity of relevant resonances for the corresponding ion and find out values of $\lambda$ for which the polarizability 
values tend to zero.

\subsection{Magnetic sublevel independent $\lambda_{\rm{magic}}$ and $\lambda_{\rm{T}}$}

We have also used our the frequency dependent scalar polarizability results of the $^{43}$Ca$^+$, $^{87}$Sr$^+$ and $^{137}$Ba$^+$ ions to
find out $\lambda_{\rm{magic}}$ that are independent of the magnetic sublevels $M_J$ of the atomic states; so also hyperfine sublevel independent. 
Table \ref{magicsd} lists of the $\lambda_{\rm{magic}}$ values for the $nS_{1/2}$-$(n-1)D_{{3/2},{5/2}}$ transitions, which lie within 
the wavelength range of 300-1500 nm. We are also able to locate the tune out wavelengths of the ground, $(n-1)D_{3/2}$ and $(n-1)D_{5/2}$ 
states of the considered alkaline ions that are independent of the $F$, $M_{F}$ and $M_{J}$ values of the respective ion and given 
them in Table \ref{tunesd}. Occurrences of these $\lambda_{\rm{magic}}$ and $\lambda_{\rm T}$ for the considered ions can offer pathways to
carry out many high precision measurements with minimal systematics.      

\section{Conclusion}

We have determined scalar, vector and tensor polarizabilities of the $nS_{1/2}$, $(n-1)D_{3/2}$ and $(n-1)D_{5/2}$ states in the $^{43}$Ca$^+$, 
Sr$^+$ and $^{137}$Ba$^+$ alkaline earth-metal ions with the ground state principal quantum number $n$. We used very precise values of the 
electric dipole matrix elements that are obtained them by employing a relativistic all-order method. 
Non-dominant contributions in the adopted sum-over-states approach for the evaluation of the polarizabilities are estimated using lower 
order perturbation methods. The obtained static polarizability values are compared with the available other theoretical results and 
experimental values to gauge their accuracies. Dynamic polarizabilities at the 1064 nm are given explicitly for the $nS_{1/2}$ and 
$(n-1)D_{3/2,5/2}$ states of the considered alkaline earth-metal ions, which could help in creating “high-field seeking“ traps using the Nd:YAG
laser. Furthermore using the dynamic polarizabilities for a wide range of wavelengths, we have located a number of tune out wavelengths 
$\lambda_{\rm{T}}$ of the above states and the magic wavelengths $\lambda_{\rm{magic}}$ for the $ \mid nS_{1/2} {F},{M_{F}}=0\rangle \rightarrow  \mid (n-1)D_{{3/2},{5/2}} {F},{M_{F}}=0\rangle$ clock transitions due to the 
circularly polarized light in the $^{43}$Ca$^+$, $^{87}$Sr$^+$ and $^{137}$Ba$^+$ alkaline earth ions. We have located a significant number
of $\lambda_{\rm{magic}}$ for these clock transitions, which can help the experimentalists to trap the above ions to reduce uncertainties 
in the clock transitions due to Stark shifts. This knowledge would also be of immense interest to carry out other high precision studies
using the considered ions. In addition, we have also determined the $\lambda_{\rm{magic}}$ and $\lambda_{\rm{T}}$ values that are independent of 
the choice of magnetic and hyperfine sublevels of the above clock transitions. 

\section{ACKNOWLEDGEMENT}

The work of B.A. is supported by Department of Science and Technology, India and work of J. K. is supported by UGC-BSR Grant No. F.7-273/2009/BSR, India. 
S.S. acknowledges the financial support from UGC-BSR. A part of the computations were carried out using Vikram-100 HPC cluster of Physical Research Laboratory and the employed SD method 
was developed in the group of Professor M. S. Safronova of the University of Delaware, USA.

%\bibliography{magicclocks2.bib}

\end{document}